\documentclass[a4paper,fleqn]{cas-dc}

\usepackage[numbers,longnamesfirst]{natbib}
\usepackage[normalem]{ulem}

\usepackage{float}

\usepackage[skip=0.333\baselineskip]{caption}
\usepackage{array}
\usepackage{subcaption}
\usepackage{adjustbox}
\usepackage{booktabs}
\usepackage{multirow}
\usepackage{makecell}
\usepackage{cleveref}

\usepackage{tabularx}
\usepackage{enumitem}

\newlist{tabitemize}{itemize}{1}
\setlist[tabitemize]{
    label=\textbullet,
    leftmargin=1.25em,
    labelsep=0.45em,
    itemsep=0pt,
    parsep=0pt,
    topsep=0pt,
    partopsep=0pt,
    before=\vspace{-0.4\baselineskip},
    after=\vspace{-0.4\baselineskip}
}

\usepackage{framed}
\usepackage{multicol}
\usepackage{nomencl}
\makenomenclature
\renewcommand*\nompreamble{\begin{multicols}{2}}
\renewcommand*\nompostamble{\end{multicols}}

\usepackage{tabularx}
\newcolumntype{Y}{>{\centering\arraybackslash}X}
\newcolumntype{Z}{>{\raggedleft\arraybackslash}X}
\usepackage{capt-of}
\usepackage{placeins}

\usepackage{enumitem}

\usepackage{tikz}
\usetikzlibrary{arrows.meta,positioning,shapes.geometric}
\usetikzlibrary{calc}
\usepackage{multirow}

\definecolor{Black}{RGB}{0, 0, 0}
\definecolor{Orange}{RGB}{230, 159, 0}
\definecolor{SkyBlue}{RGB}{86, 180, 233}
\definecolor{BluishGreen}{RGB}{0, 158, 115}
\definecolor{Yellow}{RGB}{240, 228, 66}
\definecolor{Blue}{RGB}{0, 114, 178}
\definecolor{Vermillion}{RGB}{213, 94, 0}
\definecolor{ReddishPurple}{RGB}{204, 121, 167}
\definecolor{ukpurple}{RGB}{199, 16, 92}

\usepackage{todonotes}
\newcommand{\slev}[1]{\texttt{s#1}}   
\newcommand{\old}{\texttt{old}}       
\begin{document}
\let\WriteBookmarks\relax
\def\floatpagepagefraction{1}
\def\textpagefraction{.001}

\shorttitle{AgentHomeID}
\shortauthors{H.~Ganal~et~al.}
\title[mode=title]{AgentHomeID --- Agent-based modelling of building stock transformation: A multi-scale framework for policy assessment and infrastructure planning}
\date{\today}
\author[1]{Helen Ganal}[type=editor, orcid=0009-0000-6513-5350]
\cormark[1]
\ead{helen.ganal@iee.fraunhofer.de}
\credit{Conceptualization of this study, Methodology}
\author[1]{Sarah Becker}[type=editor, orcid=0009-0002-2077-8138]
\ead{sarah.becker@iee.fraunhofer.de}
\credit{Conceptualization of this study, Methodology}
\author[1,2]{Sascha Holzhauer}[type=editor]
\ead{sascha.holzhauer@iee.fraunhofer.de}
\credit{Conceptualization of this study, Methodology}
\author[1]{Thilo Glißmann}[type=editor]
\ead{thilo.glissmann@iee.fraunhofer.de}
\credit{Conceptualization of this study, Methodology}
\author[1,2]{Friedrich Krebs}[type=editor]
\ead{friedrich.krebs@iee.fraunhofer.de}
\credit{proof-reading}
\author[1,2]{Martin Braun}[type=editor]
\ead{martin.braun@uni-kassel.de}
\credit{Supervision, proof-reading}
\author[1,2]{Philipp Härtel}[type=editor, orcid=0000-0002-9706-1007]
\ead{philipp.haertel@iee.fraunhofer.de}
\credit{Supervision, critical feedback, proofreading}
\affiliation[1]{organization={Fraunhofer IEE, Fraunhofer Institute for Energy Economics and Energy System Technology},
    addressline={Joseph-Beuys-Str. 8}, 
    city={Kassel},
    postcode={34117}, 
    country={Germany}}
\affiliation[2]{organization={University of Kassel, Sustainable Electrical Energy Systems},
    addressline={Wilhelmshöher Allee 73}, 
    city={Kassel},
    postcode={34121}, 
    country={Germany}}
\cortext[cor1]{Corresponding author}
\maketitle

\begin{abstract}
Decarbonising the building sector is central to meeting climate targets, yet existing models rarely capture the interaction between system-level transformation dynamics and heterogeneous individual investment decisions.
This work presents AgentHomeID, an agent-based model of building stock evolution in which owner behaviour, techno-economic constraints, and regulatory frameworks are represented explicitly at the level of individual buildings and their owners.
The model differentiates owner-occupiers, private landlords, and institutional owners, using willingness-to-pay (WTP) parameters estimated from empirical decision-maker studies, and operates on both representative building archetypes and real building data derived from geographic information systems (GIS).
We demonstrate this versatility across three applications.
At national scale, scenario analysis for Germany to 2045 shows that removing binding renewable heating requirements substantially raises final energy demand even where envelope refurbishment is unchanged, and that subsidy allocation and investment activity diverge sharply across owner types and income quartiles, with the lowest quartiles persistently underinvesting.
At regional scale, bottom-up simulation for a German distribution grid planning region yields spatially concentrated heat pump uptake at NUTS-3 level that differs from aggregated top-down projections in both magnitude and spatial distribution.
At urban block level, the same simulations resolve substation-level load heterogeneity and show that integrated system peaks driven by heat pumps, electric vehicles, and photovoltaics do not coincide with individual technology peaks.
Across all three scales, owner heterogeneity and local structure materially shape transition pathways, indicating that they should be represented explicitly in models used for policy assessment and infrastructure planning.
\end{abstract}
\begin{table*}[!t]
\begin{framed}
\footnotesize

\nomenclature{ABM}{Agent-based model}
\nomenclature{ABBSM}{Agent-based building stock model}
\nomenclature{ACP}{Ambitious climate protection}
\nomenclature{BAU}{Business-as-usual}
\nomenclature{BEG}{Federal Funding for Energy-Efficient Buildings}
\nomenclature{BEHG}{Fuel Emissions Trading Act}
\nomenclature{BKI}{Construction Cost Information Centre of the German Chambers of Architects}
\nomenclature{BSEM}{Building stock energy model}
\nomenclature{CO$_2$KostAufG}{CO$_2$ Cost-Sharing Law}
\nomenclature{DH}{District heating}
\nomenclature{DIN}{German Institute for Standardization}
\nomenclature{DSO}{Distribution system operator}
\nomenclature{EH 55}{Efficiency House 55}
\nomenclature{EnWG}{Energy Industry Act}
\nomenclature{EU ETS 2}{European Union Emissions Trading System 2}
\nomenclature{GEG}{Building Energy Act}
\nomenclature{GHG}{Greenhouse gas}
\nomenclature{GIS}{Geographic information system}
\nomenclature{HP}{Heat pump}
\nomenclature{IWU}{Institute for Housing and Environment}
\nomenclature{LC}{Latent class}
\nomenclature{LoD2}{Level of Detail 2}
\nomenclature{MEPS}{Minimum energy performance standards}
\nomenclature{MHP}{Municipal heat planning}
\nomenclature{NEP}{Network development plan}
\nomenclature{NPV}{Net present value}
\nomenclature{NUTS-3}{Nomenclature of Territorial Units for Statistics, level 3}
\nomenclature{ODD}{Overview, Design concepts, and Details}
\nomenclature{OFAT}{One-factor-at-a-time}
\nomenclature{SQL}{Structured Query Language}
\nomenclature{SUF}{Scientific Use File}
\nomenclature{TSO}{Transmission system operator}
\nomenclature{WPG}{Heating Planning Act}
\nomenclature{WTP}{Willingness-to-pay}

\printnomenclature[1.5cm]

\end{framed}
\end{table*}
\section{Introduction}
\label{sec:introduction}
Buildings account for 32\,\% of global energy demand and 34\,\% of energy- and process-related CO\textsubscript{2} emissions (as of 2022), with the sector's emissions rising due to population growth, expanding floor area, and greater use of energy-intensive appliances \cite{UnitedNationsEnvironmentProgrammeGlobalAllianceforBuildings.202403}.
At the European Union level, buildings account for approximately 40\,\% of final energy consumption and 36\,\% of energy-related greenhouse gas (GHG) emissions, placing the sector at the centre of the EU's climate agenda~\cite{EuropeanParliamentandtheCouncil.2024}.
At the national level, Germany's Climate Action Law requires a 65\,\% reduction in greenhouse gas emissions by 2030 compared to 1990 levels and full climate neutrality by 2045 \cite{FederalMinistryofJustice.20240715}.
Consequently, reducing emissions from the building stock is critical to meeting these policy commitments and achieving sustainable development goals \cite{IEA.02.04.2025}.

To support public policy \cite{Gilbert.2018}, simulations of the development of the building stock under various technology and regulatory scenarios have become vital tools for designing and evaluating effective mitigation strategies \cite{Hietaharju.2021,Yang.2022,Nageli.2020b, Flower.2022, Alibas.2025}.
However, the heterogeneity among buildings, including their current refurbishment status, existing heating systems, and varied ownership structures, renders projections of building stock development particularly challenging \cite{Cayla.2015}.
Decisions made at the level of building owners influence both the building’s energy performance and the chosen energy carrier, but need to respect which energy carriers the infrastructure supplies \cite{Dodds.2014, Michelsen.2012, Henkel.2012, Bauermann.2014, Bauermann.2015}.
Furthermore, the ongoing decarbonisation transition, shifting from fossil-fuel-based heating (predominantly natural gas and oil) towards electricity-driven heat pumps and district heating, limits the applicability of historical experience \cite{Wilson.2018}.

Building stock energy models (BSEMs) tackle the outlined challenges. BSEMs are computational simulation tools designed to evaluate energy consumption patterns and environmental impacts across building portfolios and to identify pathways to reduce both energy demand and associated greenhouse gas emissions. As defined by \cite{Nageli.2022}, BSEMs are characterised by three core features:
\begin{enumerate}[label=\alph*)]
    \item \textit{Building heterogeneity}: They typically model multiple buildings, which may or may not be geographically clustered.
    \item \textit{(Energy) indicator output}: They quantify performance through measurable indicators such as final energy consumption or carbon intensity.
    \item \textit{Predictive capability}: They generate projections beyond observed data ranges to simulate different scenario conditions.
\end{enumerate}
We argue that for credible and reliable policy decision support, all three features require particular attention.
From them, we derive three modelling requirements: the first addresses the data foundation of building heterogeneity, the second the behavioural mechanisms underlying predictive capability, and the third the boundary conditions under which both operate.
The indicator outputs, in turn, link the resulting stock dynamics to policy-relevant metrics.

Accurately capturing building heterogeneity necessitates the representation of diverse building typologies and attributes, encompassing current refurbishment status, installed heating technologies, and ownership structures, with a high degree of specificity.
BSEMs must incorporate \textbf{(1) transparent data interfaces} capable of accessing and integrating the growing volume of digital building inventory data, such as digital twins of the built environment, sourced from building registries, energy performance certificates, and related databases.

Similarly, BSEMs should couple behavioural dynamics and stock evolution to a coherent suite of energy indicator outputs.
This approach enables the simulation of final energy demand and carbon intensity trajectories, thereby generating policy-relevant metrics, including sectoral emission reductions, renovation rates, and technology diffusion pathways, directly from model outputs.

To enable predictive capabilities, BSEMs must account for the mechanisms influencing building stock dynamics over time.
Structural modifications to buildings primarily arise from refurbishment activities undertaken by respective owners.
To adequately represent this complexity, \textbf{(2) the heterogeneity of building owners}, including typologies, socio-economic characteristics, and preferences, should be systematically incorporated into BSEMs based on robust empirical evidence.

These highly decentralised drivers and aggregated indicators of building stock development are complemented by overarching scenario-defined boundary conditions, which feed back to the building owner level and critically influence simulation outcomes.
To provide policy support, BSEMs need to integrate \textbf{(3) close-to-reality, comprehensive representations of legislative frameworks and assumptions on future energy infrastructures, both existing and to be designed}.

In this work, we present \textsc{AgentHomeID} (Agent-based modelling of the building stock based on Homeowners Investment Decisions), a building stock energy model that comprehensively addresses the requirements outlined above.
With respect to \textbf{(1) transparent data interfaces}, \textsc{AgentHomeID} can be operated with both a representative synthetic building stock derived from the German micro census and detailed regional datasets based on publicly available building geometry data, enabling consistent analysis across spatial scales.

With respect to \textbf{(2) the heterogeneity of building owners}, \textsc{AgentHomeID} distinguishes private owner-occupiers, landlords, and owners' associations as well as institutional owners, each characterised by ownership-type-specific socio-economic properties and empirically derived heterogeneous preferences.
With respect to \textbf{(3) legislative frameworks and energy infrastructure}, \textsc{AgentHomeID} integrates the current German regulatory environment, alongside infrastructure-dependent emission factors for electricity, district heating, and gas mixtures. Finally, \textsc{AgentHomeID} supports multi-scale applications from national policy assessment to municipal heat planning, supporting analysis of technology diffusion, distributional impacts, and spatially resolved infrastructure needs to meet building sector decarbonisation.

The novel contribution of this paper lies in providing a comprehensive methodological description of the AgentHomeID framework and, by consolidating previously published application contexts across spatial scales, synthesising its capabilities into a unified assessment of the model's scope for policy and infrastructure applications.

The remainder is organised as follows: Section~\ref{sec:previous} positions \textsc{AgentHomeID} within the existing landscape of building stock energy models, reviews related agent-based approaches, and identifies the specific research gaps that \textsc{AgentHomeID} addresses.
Section~\ref{sec:methods} details the model structure and input data of \textsc{AgentHomeID}, covering the building stock representation, owner agent typology and decision-making processes, regulatory framework integration, energy and cost calculations, and spatio-temporal resolution.
Section~\ref{sec:applications} demonstrates the model's capabilities through a set of real-world application examples spanning national policy assessment, regional distribution grid planning, and municipal heat planning.
Section~\ref{sec:discussion} discusses the results in the context of the identified research gaps and reflects on model limitations and avenues for future work.
Section~\ref{sec:conclusion} concludes the paper.
\section{Previous work}
\label{sec:previous}

\begin{figure*}[pos=htbp]
    \centering
    \includegraphics[width=1\linewidth]{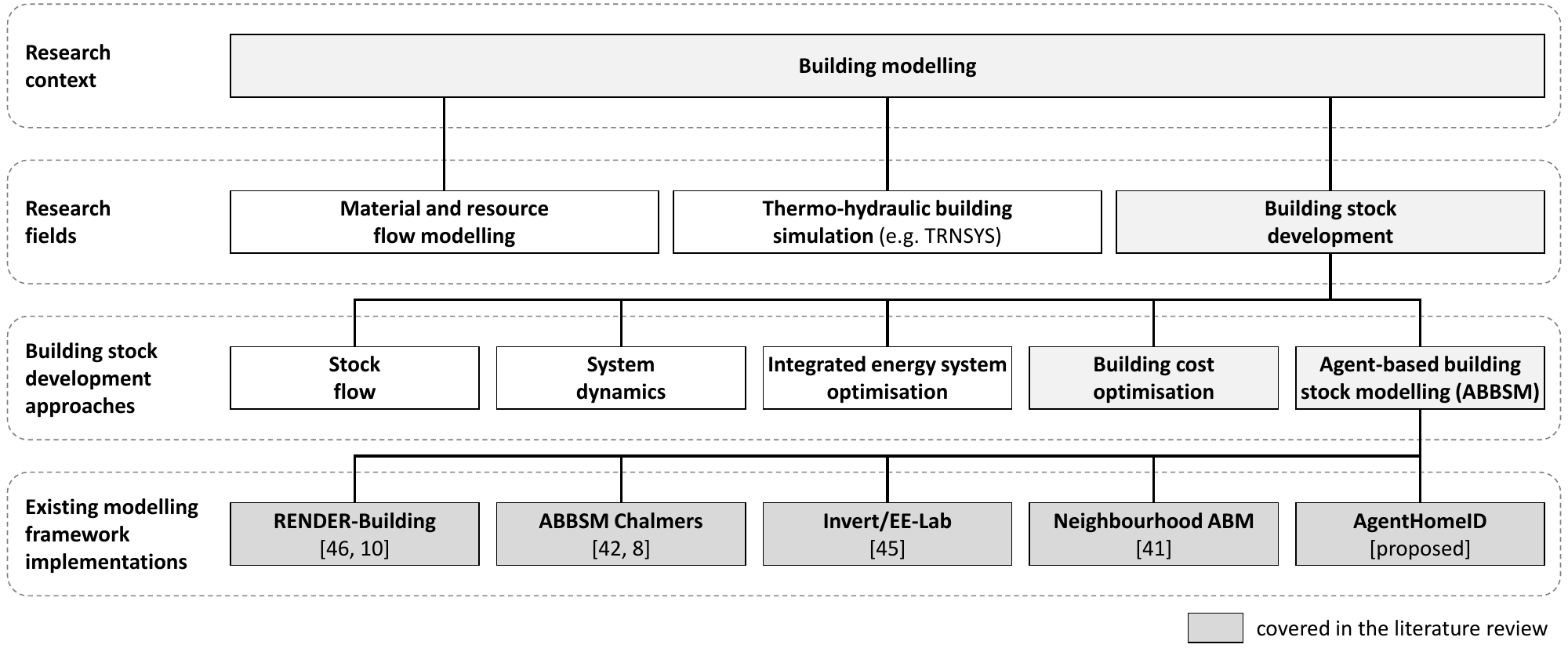}
    \caption{Mapping comparable research fields, building stock development approaches, and existing model implementations; scope delineation of this work.}
    \label{fig:model_delimitation}
\end{figure*}

\paragraph{Building stock model landscape and delimitation}
The modelling of building sector transformation spans several related but distinct research fields operating at different levels of detail and abstraction (see Fig.~\ref{fig:model_delimitation}). Highly resolved building simulation environments such as \textsc{TRNSYS} enable detailed thermo-physical analyses at the individual building level but are not designed to represent long-term stock evolution at regional or national scale \cite{Ibanez.2005, Tashtoush.2015}.
Material and resource flow models focus on embodied emissions, material cycles, and construction dynamics \cite{Mostert.2022, Knoeri.2013}, while integrated energy system models typically optimise technology deployment and infrastructure development from a system perspective.
\par

BSEMs occupy an intermediate position by representing the evolution of building stocks and their energy demand over extended time horizons \cite{Swan.2009, Langevin.2020}.
According to the classification framework proposed by \cite{Langevin.2020}, BSEMs can be differentiated by their treatment of energy systems, people, building stocks, and environmental context.
Within this landscape, \textsc{AgentHomeID} is positioned at the intersection of dynamic building stock modelling, owner decision modelling, and policy-sensitive agent-based simulation.
The following review is structured around three key modelling requirements introduced in Section~\ref{sec:introduction}: the representation of building heterogeneity, owner heterogeneity, and regulatory and infrastructure frameworks.

\paragraph{Building heterogeneity and building stock representation}
Building stock energy models have become established tools for analysing long-term energy demand, technology diffusion, and emissions in the building sector \cite{Swan.2009, Langevin.2020}.
Many existing approaches represent building stocks through archetypes or aggregated stock segments and are particularly strong in stock accounting, scenario analysis, and techno-economic projections.
Examples include national and regional models developed for Germany and Europe \cite{McKenna.2013, Mata.2014, Sandberg.2016, Senkpiel.2020, Sandberg.2017, Sandberg.2021, Yamaguchi.2022}. These models capture construction, demolition, refurbishment, and energy demand dynamics, but generally represent renovation behaviour and technology adoption in a stylised manner.

Recent multi-model comparisons further demonstrate that building-sector
transformation pathways can depend substantially on model structure.
\cite{Ozer.2026} compare five structurally different building-sector energy
models under harmonised EU27 decarbonisation scenarios and identify
electrification and building renovation as comparatively robust outcomes,
whereas district-heating deployment and the pace of technology diffusion
vary more strongly across models. These differences are attributed
particularly to decision-making logic, spatial resolution, renovation
dynamics, and technology-diffusion mechanisms. 

More recent developments have focused on improving the representation of building heterogeneity.
Synthetic stock generation approaches provide a bridge between aggregate stock accounting and explicit building representation.
In particular, \cite{Nageli.2018} demonstrate how synthetic building stocks can preserve substantial heterogeneity beyond conventional archetype approaches, while \cite{Aldenhoff.2024} highlight the importance of refurbishment cycles for realistic retrofit projections.

Nevertheless, many existing BSEMs still rely on simplified representations of ownership structures and decision-making processes.

\paragraph{Owner heterogeneity and decision-making}
A second strand of literature focuses on the behavioural processes underlying refurbishment and technology adoption decisions.
In Germany, discrete-choice and simulation studies have investigated heating technology adoption, policy incentives, and path dependencies in the residential heat market \cite{Michelsen.2012, Henkel.2012, Bauermann.2014, Bauermann.2015}.
These studies consistently show that investment decisions are influenced not only by economic considerations but also by infrastructure availability, existing system configurations, environmental preferences, and policy design.

Reviews of renovation and technology adoption modelling indicate that empirically grounded representations of owner behaviour remain comparatively scarce, particularly when multiple technologies, socio-economic characteristics, building properties, and regulatory constraints are considered simultaneously \cite{Friege.2014, Du.2022, Hesselink.2019}.
For the German context, \cite{Friege.2016} is especially relevant, as it models insulation decisions based on survey-derived behavioural rules, situational triggers, and social interactions, thereby demonstrating the importance of non-financial decision drivers.

Agent-based approaches provide a natural framework for representing heterogeneous actors and boundedly rational decision-making.
International applications include studies of heat pump adoption in Ireland \cite{Meles.2022}, policy interventions in the United Kingdom \cite{Flower.2022}, heating and cooling technology diffusion \cite{Du.2024}, and combined heating system replacement and envelope refurbishment decisions in the Netherlands \cite{Akhatova.2025}.
These studies demonstrate the ability of agent-based models to capture behavioural heterogeneity, policy responses, and non-linear diffusion dynamics that are difficult to represent in aggregate modelling approaches.

Among the most influential contributions are the Swiss agent-based building stock models (ABBSMs) developed at Chalmers by \cite{Nageli.2020, Nageli.2020b}, which combine synthetic building stock generation, endogenous refurbishment and heating system decisions, bounded rationality, and policy evaluation.
More recently, \cite{Alibas.2025} developed RENDER-Building, a German national-scale agent-based building stock model with high spatial resolution, spatially differentiated infrastructure availability, and a broad policy portfolio.

While these approaches significantly advance the representation of owner decision-making, comparatively less attention has been devoted to detailed ownership differentiation, landlord--tenant constellations, and the interaction of heterogeneous ownership structures with distribution-sensitive policy instruments.

\paragraph{Regulatory frameworks, infrastructure constraints, and distributional effects}
Building stock transformation is strongly influenced by regulatory frameworks and the availability of energy infrastructures. Existing modelling approaches increasingly incorporate policy instruments such as subsidies, carbon pricing, technology bans, and renovation obligations \cite{Nageli.2020b, Flower.2022, Alibas.2025}. However, the representation of regulatory interactions often remains focused on aggregate technology adoption and energy system impacts.

A growing body of literature highlights the importance of ownership structures and landlord--tenant relationships for evaluating policy effectiveness. \cite{George.2023} analyse interactions between carbon pricing, redistribution mechanisms, and modernisation incentives, while \cite{Reutter.2025} investigate retrofit incentives under German tenancy law and detailed cost-allocation rules. These studies demonstrate that policy evaluation in the building sector cannot be reduced to average cost-effectiveness indicators alone, but must also consider distributional impacts across different owner and tenant groups.

Furthermore, several recent studies emphasise the growing importance of infrastructure availability for technology adoption decisions, particularly regarding district heating expansion, hydrogen infrastructure, and electricity network development \cite{Akhatova.2025, Alibas.2025}. Integrating infrastructure constraints and policy frameworks consistently with heterogeneous owner decisions therefore remains an important challenge for building stock modelling.

\paragraph{Research gap and positioning of \textsc{AgentHomeID}}
The reviewed literature demonstrates substantial progress in representing building stock dynamics, owner decision-making, and policy interventions.
However, existing approaches typically address only a subset of the three key modelling requirements identified in Section~\ref{sec:introduction}.
Models with detailed building stock representations often simplify owner behaviour, and even the behaviourally rich, infrastructure-aware approaches \cite{Nageli.2020b, Alibas.2025} represent owners as a homogeneous or only coarsely differentiated investor class, leaving landlord--tenant constellations, different decision-maker types and the distributional incidence of policy instruments outside the model scope. 

\textsc{AgentHomeID} addresses these gaps by combining (i) a detailed representation of building heterogeneity, (ii) empirically grounded owner heterogeneity with differentiated ownership structures and behavioural preferences, and (iii) explicit modelling of regulatory frameworks and infrastructure-dependent transformation pathways within a unified agent-based framework.
In addition, the model supports applications across multiple spatial scales, ranging from national policy assessment to regional infrastructure planning and municipal heat planning.

Compared with the most closely related approaches \cite{Nageli.2020, Nageli.2020b, Kranzl.2013, Akhatova.2025, Alibas.2025}, \textsc{AgentHomeID} places particular emphasis on ownership differentiation, distribution-sensitive policy analysis, and the interaction between decentralised building-level decisions and infrastructure development.

A detailed feature comparison of the most closely related modelling approaches is provided in Tables~\ref{tab:model_comparison_a} and~\ref{tab:model_comparison_b}. The comparison is based on explicitly documented model capabilities rather than on model labels alone and distinguishes between the representation of decision units, behavioural mechanisms, ownership structures, spatial and technical resolution, infrastructure, policy, and the granularity of energy, emissions, and cost calculations. The decisive differences concern owner differentiation (landlord vs.\ owner-occupier), empirically estimated willingness-to-pay in the decision model, and the option to operate on GIS-based real building geometries. Conversely, \textsc{AgentHomeID} does not resolve hourly energy-demand profiles, which RENDER-Building provides. A ``+'' denotes an explicitly and substantially represented feature, ``(+)'' a limited or partial representation, ``(-)'' an indirect, aggregated, or exogenous representation, and ``-'' indicates that the feature is not represented in the referenced model description. The assigned ratings refer to the capabilities documented in the cited publications and do not necessarily imply an identical degree of detail, modelling philosophy, or underlying data resolution across the compared approaches.

\begin{table*}[t]
\caption{Comparison of agent-based and closely related bottom-up building stock transformation models across key representation domains (Part a).}
\centering
\scriptsize
\renewcommand{\arraystretch}{1.12}
\setlength{\tabcolsep}{2pt}

\begin{adjustbox}{width=\textwidth}
\begin{tabular}{
p{2.65cm}
p{4.35cm}
p{1.55cm}
p{1.55cm}
p{1.55cm}
p{1.55cm}
p{1.55cm}
}
\toprule
\textbf{Representation domain}
& \textbf{Sub-aspects}
& \textbf{RENDER-Building}\newline Fh ISI
& \textbf{ABBSM}\newline Chalmers
& \textbf{Invert/EE-Lab}\newline TU Wien
& \textbf{Neighbour-hood ABM}\newline TU Wien
& \textbf{AgentHomeID}\newline Fh IEE \\
&
& \cite{Alibas.2024,Alibas.2025}
& \cite{Nageli.2020,Nageli.2020b}
& \cite{Kranzl.2013}
& \cite{Akhatova.2025}
& proposed \\
\midrule

\multirow{2}{*}{\textbf{\makecell[l]{Model architecture\\
\& decision unit}}}
& Building-level decision units
& + & + & (-) & + & + \\

& Decision maker explicitly represented
& (-) & (+) & (-) & + & + \\
\midrule

\multirow{6}{*}{\textbf{\makecell[l]{Decision-making\\
\& behavioural\\
representation}}}
& Event-triggered investment decisions
& + & + & (+) & + & + \\

& Bounded-rational or probabilistic option choice
& + & + & + & (+) & + \\

& Socio-demographic attributes directly affect investment decisions
& - & - & (-) & - & + \\

& WTP term included in option utility
& - & + & + & - & + \\

& WTP informed by empirical decision-maker studies
& - & (+) & (-) & - & + \\

& Owner-role-specific WTP estimated specifically for the model
& - & - & - & - & + \\
\midrule

\multirow{4}{*}{\textbf{\makecell[l]{Decision-maker\\
\& ownership\\
differentiation}}}
& Private and institutional owner types
& - & - & (-) & - & + \\

& Owner-occupier and landlord differentiation
& - & - & (-) & - & + \\

& Ownership-specific decision rules or parameters
& - & (-) & (+) & - & + \\

& Tenant--landlord relationship and cost allocation
& - & - & - & - & + \\
\midrule

\multirow{6}{*}{\textbf{\makecell[l]{Spatial scope\\
\& stock\\
representation}}}
& European or multi-country applications
& - & - & + & - & - \\

& National-scale application
& + & + & + & - & + \\

& Regional or district-scale application
& + & (-) & + & (-) & + \\

& Neighbourhood or urban-block application
& (-) & - & (-) & + & + \\

& Spatially located building agents
& + & (-) & (-) & - & + \\

& Explicit real-building and GIS-based input supported
& - & - & - & - & + \\
\midrule

\multirow{3}{*}{\textbf{\makecell[l]{Temporal scope\\
\& stock dynamics}}}
& Long-term scenario horizon
& + & + & + & + & + \\

& Annual or multi-year simulation steps
& + & + & + & + & + \\

& Endogenous ageing, refurbishment, and stock turnover
& + & + & + & (-) & + \\
\midrule

\multirow{5}{*}{\textbf{\makecell[l]{Building\\
\& technology\\
representation}}}
& Residential building stock
& + & + & + & + & + \\

& Non-residential building stock
& + & - & + & - & + \\

& Envelope components represented separately
& + & + & + & (+) & + \\

& Multiple heating technologies and energy carriers
& + & + & + & (-) & + \\

& Coupled envelope and heating-system choices
& (+) & (+) & (+) & + & + \\
\midrule

\multirow{5}{*}{\textbf{\makecell[l]{Infrastructure\\
\& location-specific\\
feasibility}}}
& Spatially differentiated infrastructure availability
& + & + & (+) & - & + \\

& Building-level infrastructure connection status
& (+) & + & (-) & - & + \\

& Building-specific renewable-heat feasibility
& (-) & + & (+) & - & + \\

& Explicit network topology or capacity representation
& - & - & - & - & (+) \\

& Infrastructure evolution and municipal heat-planning integration
& (+) & (+) & (-) & - & + \\

\bottomrule
\end{tabular}
\end{adjustbox}

\begin{minipage}{\textwidth}
\footnotesize
\textit{Note:}
A ``+'' denotes an explicitly and substantially represented feature,
``(+)'' a limited or partial representation,
``(-)'' an indirect, aggregated, or exogenous representation,
and ``-'' indicates that the feature is not represented in the referenced model description.
\end{minipage}

\label{tab:model_comparison_a}
\end{table*}

\begin{table*}[t]
\caption{Comparison of agent-based and closely related bottom-up building stock transformation models across key representation domains (Part b).}
\centering
\scriptsize
\renewcommand{\arraystretch}{1.12}
\setlength{\tabcolsep}{2pt}

\begin{adjustbox}{width=\textwidth}
\begin{tabular}{
p{2.65cm}
p{4.35cm}
p{1.55cm}
p{1.55cm}
p{1.55cm}
p{1.55cm}
p{1.55cm}
}
\toprule
\textbf{Representation domain}
& \textbf{Sub-aspects}
& \textbf{RENDER-Building}\newline Fh ISI
& \textbf{ABBSM}\newline Chalmers
& \textbf{Invert/EE-Lab}\newline TU Wien
& \textbf{Neighbourhood ABM}\newline TU Wien
& \textbf{AgentHomeID}\newline Fh IEE \\
&
& \cite{Alibas.2024,Alibas.2025}
& \cite{Nageli.2020,Nageli.2020b}
& \cite{Kranzl.2013}
& \cite{Akhatova.2025}
& proposed \\
\midrule

\multirow{5}{*}{\textbf{\makecell[l]{Policy\\
\& regulatory\\
representation}}}
& Carbon pricing represented in operating costs
& + & + & (+) & - & + \\

& Investment subsidies or operating support
& + & + & + & + & + \\

& Technology bans, requirements, or efficiency standards
& + & + & + & + & + \\

& Detailed country-specific legal interactions
& (-) & + & (+) & (+) & + \\

& Finite policy budgets or resource constraints
& (-) & - & (+) & - & + \\
\midrule

\multirow{8}{*}{\textbf{\makecell[l]{Energy, emissions\\
\& cost\\
assessment}}}
& Useful heat demand per modelled building unit
& + & + & (-) & + & + \\

& Final energy demand per modelled building unit
& + & + & (-) & + & + \\

& Norm-based design heating load
& (-) & (+) & - & (-) & + \\

& Hourly demand profiles generated endogenously
& + & - & - & - & - \\

& Post-processed load profiles or peak-load indicators
& + & - & - & - & + \\

& GHG emissions calculated per modelled building unit
& + & + & (-) & + & + \\

& Investment and operating costs per decision unit
& + & + & (-) & + & + \\

& Distributional results by ownership or income group
& - & - & - & - & + \\

\bottomrule
\end{tabular}
\end{adjustbox}

\begin{minipage}{\textwidth}
\footnotesize
\textit{Note:} See Table~\ref{tab:model_comparison_a} for the rating scheme.
\end{minipage}

\label{tab:model_comparison_b}
\end{table*}
\section{\textsc{\textsc{AgentHomeID}} modelling framework}
\label{sec:methods}
We describe our framework following the best-practice reporting guideline for building stock energy models presented in \cite{Nageli.2022} and considering relevant parts of the Overview, Design concepts, and Details (ODD) protocol~\cite{Grimm.2020} that address agent-based models.
We explicitly report the design concepts of agents' objectives, sensing, and prediction in the People subsection, and consider stochasticity in the Data processing subsection.

\subsection{Overview}
\subsubsection{Aim and scope}
\textsc{AgentHomeID} is an agent-based model designed to simulate investment decisions by building owners and project the evolution of the building stock over time.
The model is developed to support a broad range of application domains, spanning analyses at national, regional, and local levels, including policy assessment, infrastructure planning, and municipal heat planning.
Depending on the application context, simulations can be conducted either using a representative building stock derived from unlocalised sample data or based on detailed regional datasets.
Both data approaches are applied in this study to demonstrate the versatility of the modelling framework across different spatial scales and planning contexts.
The model environment represents a regional building stock in which individual agents (i.e., building owners) interact with external influences such as regulatory changes, cost trends, and fuel price dynamics. 
\subsubsection{Modelling approach}
\begin{figure}[h]
\centering
\begin{tikzpicture}[
  scale=1,
  transform shape,
  font=\small,
  node distance=3.5mm and 10mm,
  >=Latex,
  block/.style={rectangle, draw, rounded corners=2pt, align=center,
                minimum width=50mm, minimum height=4.8mm},
  decision/.style={diamond, draw, aspect=3.8, align=center,
                   inner sep=1pt, minimum width=50mm},
  note/.style={font=\footnotesize, align=center, text width=52mm}
]

\node[block] (bldg) {Consider owned building};

\node[decision, below=of bldg] (trigger) {Refurbishment trigger?};

\node[decision, below=of trigger] (cancel) {Cancel refurbishment?};

\node[block, below=of cancel] (gen) {Generate refurbishment options};

\node[block, below=of gen] (constraints) {Apply technical, financial\\and infrastructure constraints};

\node[block, below=of constraints] (choose) {Select refurbishment option};

\node[block, below=of choose] (exec) {Implement refurbishment};

\node[block, below=of exec] (done) {Done};

\draw[->] (bldg) -- (trigger);
\draw[->] (trigger) -- node[right, font=\footnotesize]{Yes} (cancel);
\draw[->] (cancel) -- node[right, font=\footnotesize]{No} (gen);
\draw[->] (gen) -- (constraints);
\draw[->] (constraints) -- (choose);
\draw[->] (choose) -- (exec);
\draw[->] (exec) -- (done);

\draw[->] (trigger.east) -- ++(10mm,0) node[above, font=\footnotesize]{No} |- (done.east);

\draw[->] (cancel.east) -- node[near start, right, yshift=2mm, font=\footnotesize]{Yes} ++(10mm,0) |- (done.east);

\end{tikzpicture}
\caption{Refurbishment decision workflow executed for each owner agent at every model time step.}
\label{fig:fc_overview_tikz}
\end{figure}
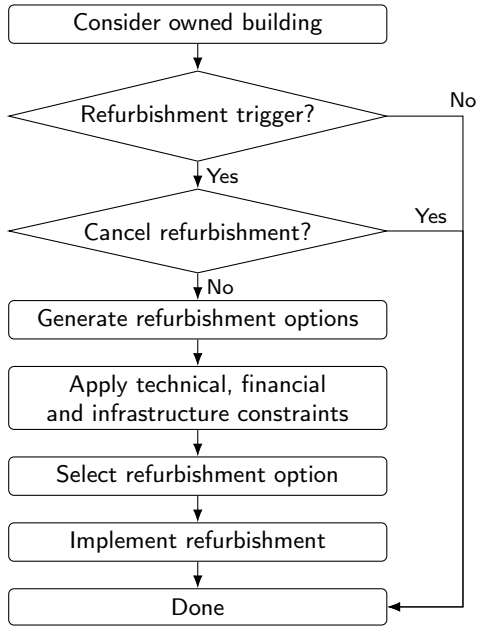

As an agent-based building stock model, \textsc{AgentHomeID} follows a bottom-up approach with a high degree of transparency in modelling building owners' decision-making (white box approach).
Main modelling parts are buildings, including their physics and refurbishment states of particular building parts and their associated owners who decide about refurbishment measures in a way that depends on their ownership type.
In each time step, each owner agent is activated once, checks its building, and
potentially refurbishes it, as depicted in Fig.~\ref{fig:fc_overview_tikz}.

This process consists of the following steps: Firstly, the agent checks itself as well as the owned building for refurbishment triggers, such as a change in ownership or a building part reaching the end of its lifetime.
If such a trigger is not present, the agent finishes for this time step.
If there is one or several triggers present, a probabilistic cancellation is applied to represent model exogenous factors such as a lack of workforce.
If the refurbishment is cancelled, the building remains unchanged.
Otherwise, a list of refurbishment options is compiled.
These options are then filtered for their internal consistency (e.g., the installation of a heat pump requires a sufficiently well-insulated building envelope), and the financial viability for the owner.
The resulting list of refurbishment options is then passed to the agents' decision process, and one option is chosen.
The chosen refurbishment is executed, i.e. the building state is updated accordingly.

The contents of the following subsections and their sources are summarised in Table \ref{tab:data_sources} of the Appendix.
\subsubsection{System boundaries}
The simulation represents individual residential buildings as explicit agents in both of its application modes.
The two modes differ only in how the agent population is constituted.
In the national mode, a representative sample of approximately 40{,}000 buildings from the German micro census scientific use file (SUF) \cite{SUF} is simulated explicitly and mapped to the full German stock through building-specific representativity weights.
In the regional mode, all buildings of a study area are simulated directly from local data, in which case all representativity weights equal one.
In both modes, every decision, energy, and cost calculation is carried out at the level of the individual building agent rather than on aggregated stock segments; the weights affect only the aggregation of results, not the decision logic.

Non-residential buildings are represented analogously. In the national mode, they are statistically initialised according to the distribution of building function types and construction age classes. In the regional mode, local building information is used to assign individual buildings to the corresponding Institut Wohnen und Umwelt (IWU) building function classes and to determine their construction age. The simulated energy services comprise space heating, domestic hot water, and ventilation.

The model supports a broad range of applications across different spatial scales and planning contexts. The supported application domains and exemplary use cases are presented in more detail in Section~\ref{sec:applications}.

\subsubsection{Spatio-temporal resolution }
The simulation period covers the years 2020 to 2045 with a five-year time step, reflecting the slow evolution and long-term dynamics of the building stock. The spatial resolution is per building. 
\subsection{Model components}
\label{subsec:model_components}
\subsubsection{Building stock}
\label{sec:building_stock}
The model simulates a set of individual residential buildings of various dwelling types which can be scaled according to their representation factor. The initial residential building stock is statistically derived from the German SUF~\cite{SUF}. Building characteristics are assigned based on the IWU residential building typology \cite{iwu:2016} and representative residential archetypes that were derived from Germany-wide Level of Detail 2 (LoD2) building data. Additional adjustments are applied to better represent the national building stock. Non-residential buildings are represented by statistically sampled archetypes characterised by type of usage and function (office, education, leisure, sports, catering and accommodation, production and logistics, and trade) and age class \cite{Horner.2022}.

Alternatively, it is possible to use locally or regionally available data and derive building characteristics with the methods presented in \cite{horst_2025_ut}. A similar approach to derive part of the required input data from public data sets has been reported in \cite{Dabrock.2025}.

Each building is characterised by envelope parts (roof, exterior walls, windows, and the lower parts) and their condition (age, refurbishment state, heat transfer coefficients (\textit{k}-values)), technical system condition (heating system type; the presence or absence of ventilation systems with heat recovery; energy carrier; and features such as night-time temperature setback), usage area, and construction year.

\subsubsection{People}
\label{sec:people}
Building owners are represented as autonomous individual agents. For residential buildings, owners are categorised into two main groups: private individuals and institutional owners. Private individuals are further segmented into owner-occupiers, private landlords, and homeowner associations. Institutional owners comprise private property companies, cooperatives, and public authorities based on statistical data from the Federal Statistical Office (Statistisches Bundesamt).

For non-residential buildings, ownership information is generally unavailable in the underlying datasets. Therefore, ownership types are assigned probabilistically according to the building function class based on statistical assumptions. Accordingly, non-residential buildings are assigned exclusively to the institutional ownership categories of private property companies, cooperatives, and public authorities.

This categorisation enables the model to capture heterogeneous ownership structures and decision behaviours in response to economic and regulatory stimuli while accounting for the different data availability in the residential and non-residential sectors.

Agents are assumed to possess accurate information regarding refurbishment costs and to be fully aware of all available funding schemes.
Private owners consider current energy prices without forecasting future price developments.
Institutional owners use NPV-based economic evaluations.
For self-use-oriented decisions, applied to cooperatives and selected public authorities, future fuel-cost trajectories are considered.

Decisions of private individuals are driven by a personal random utility valuation.
The latent-class (LC) assignment is based on socio-economic, building-related,
and behavioural variables comprising age, income, education, gender,
municipality size, building size, previous modernisation activities,
construction year, and building condition. Separate class-membership
models are used for private owner-occupiers and private landlords, each
comprising three latent classes. For each agent, class-membership
probabilities are calculated from the corresponding assignment variables,
class-specific coefficients, and a class-specific constant, and one latent
class is subsequently drawn according to these probabilities. The statistical significance of the individual assignment variables varies across latent classes.

Each latent class is associated with a distinct set of WTP values for the attributes considered in the utility calculation,
based on empirical data obtained from a discrete choice experiment
\cite{Bender2026}.

The WTP-based utility specification considers investment costs, subsidies,
energy savings, CO$_2$ reductions, unavoidable investment requirements,
changes in operating costs, heating-system type, envelope insulation level,
and the decision whether to refurbish at all.

The numerical WTP estimates and latent-class assignment coefficients cannot
be disclosed in the present paper due to contractual restrictions associated
with the underlying empirical study. To enable assessment of the
decision-model structure, the functional form, the variables entering the
latent-class assignment, and the attributes entering the utility calculation
are therefore reported explicitly in this section.

The decision-making process is myopic.
However, key characteristics of refurbishment measures (as identified through consultation with an energy advisor) are assumed to be known.

In \textsc{AgentHomeID}, the WTP value assignment of the agents representing private individuals proceeds in two steps: First, for each agent and each LC, the probability of the agent belonging to the LC based on the corresponding assignment variables is calculated as follows:
\begin{align}
    p_{i,q} = \frac{\exp\left(\beta_{q,0} + \sum_n \beta_{q,n} x_{i,n}\right)}{\sum_{q'} \exp\left(\beta_{q',0} + \sum_n \beta_{q',n} x_{i,n}\right)}
    \label{eq:lc_prob}
\end{align}
where $p_{i,q}$ is the probability of agent $i$ belonging to latent class $q$, $\beta_{q,0}$ is the class-specific constant, $\beta_{q,n}$ is the coefficient of assignment variable $n$ for latent class $q$, and $x_{i,n}$ is the realisation of assignment variable $n$ for agent $i$.
The denominator sums over all latent classes and ensures normalisation, such that $\sum_{q'} p_{i,q'} = 1$.

The effect of the $\beta$ parameters can be illustrated using age as an example: $x_{i,\mathrm{age}}$ denotes the age of agent $i$ in years.
For an LC with positive $\beta_{q, \mathrm{age}}$, increasing age will increase the probability of belonging to LC $q$. If $\beta_{q, \mathrm{age}}$ is negative, increasing age will decrease the probability of belonging to this LC.
Based on the probability distribution given by \eqref{eq:lc_prob}, each agent is randomly assigned to exactly one latent class using a weighted random draw. This assignment is performed once during model initialisation and remains fixed throughout the simulation. In the second step, the empirically derived WTP values corresponding to the assigned latent class are ascribed to the agent.

Decisions of institutional owners are based on a net present value (NPV)
assessment of the available investment options, with the evaluation approach
depending on the ownership role (Table~\ref{tab:owner_decision_logic}).
For self-use-oriented decisions, applied to cooperatives and selected public
authorities, avoided fuel-cost savings are discounted over the technical
lifetime of the respective building component or heating system.
For rental-oriented decisions, applied to private property companies and
selected public authorities, rental earnings and the owner's share of
CO$_2$ costs are evaluated over an owner-specific maximum payback period.
Investment costs and available subsidies are considered in both approaches,
and owner- and trigger-specific rates of return are used as discount rates.
The corresponding parameter assumptions are summarised in
Table~\ref{tab:npv_assumptions}.

\subsubsection{Environment}
\label{sec:environment}
\paragraph{Regulatory framework}
The model explicitly integrates external policy signals. The regulatory framework implemented in the present model version reflects the legislation and funding conditions in force as of May 2026 and comprises four key instruments:
\begin{itemize}
    \item \textit{Building Energy Act (GEG – 09/2023)} \cite{GEG}, which requires new heating systems to use at least 65\,\% renewable energy, mandates the Efficiency House 55 (EH 55) standard for new buildings, and integrates municipal heating planning;
    \item \textit{Heating Planning Act (WPG – 12/2023)} \cite{WPG}, which sets renewable energy and waste heat targets for heating networks (2030/2045) and mandates municipal heating plans (MHP) by June 2026 (cities $>$100,000 inhabitants) and June 2028 (smaller municipalities);
    \item \textit{Fuel Emissions Trading Act (BEHG – 12/2019, revised 12/2023)} \cite{ETS2, BEHG}, which introduces greenhouse gas pricing for fuels, rising from 25\,€/t to 55\,€/t CO\textsubscript{2} by 2025, with a transition to the European Union Emissions Trading System 2 (EU ETS 2) in 2027; and
    \item \textit{Federal Funding for Energy-Efficient Buildings (BEG – 12/2023)} \cite{BEG}, which supports full-scale renovations, renewable heating subsidies, climate bonuses, and income-based incentives for owner-occupiers, with extra aid for worst-performing buildings and phased renovations.\newline
\end{itemize}
Additionally, the $\mathrm{CO_2}$ Cost-Sharing Law (CO2KostAufG) is accounted for, which governs the cost sharing of emission certificates between landlords and tenants.

Further policy instruments beyond the current regulations can be selected depending on the scenario.
These instruments may be introduced and phased out at flexible points in time and can be combined with one another.
These policies influence both the frequency and the type/extent of renovation decisions, thereby affecting the overall pace of decarbonisation in the heating sector.
The set of optional intervention instruments includes:
\begin{itemize}[noitemsep, topsep=0pt]
    \item gas grid decommissioning pathways
    \item budget constraints (e.g.\ for biomass, funding volumes, hydrogen, and biomethane)
    \item constraints on the share of income and the time horizon over which investment volumes are determined
    \item limits on the permissible lifetime of fossil heating systems
    \item limits on the permissible lifetime of non-refurbished building components
    \item introduction of minimum energy performance standards (MEPS)
    \item minimum efficiency standards for new buildings
    \item accounting-based use of biomethane and hydrogen
    \item provision of information on future price developments
    \item alternative funding schemes to the current subsidy framework, including income-dependent and efficiency-based funding rates
\end{itemize}
\paragraph{Spatial environment}
Regarding the physical context, several building-specific spatial characteristics are considered.
In addition to building orientation, which is used to calculate solar energy input, geometric properties such as component surface areas and building volumes are represented.
In addition, building-specific infrastructure connection possibilities are considered, including access to district heating, hydrogen, and gas networks.
Spatial characteristics are further used to determine technology potentials, such as the share of green areas as an indicator for the feasibility of ground-source heat pumps and distances to neighbouring buildings for assessing installation constraints of air-source heat pumps.

\subsubsection{Energy}
\begin{figure}[pos=htbp]
    \centering
    \includegraphics[width=1\linewidth]{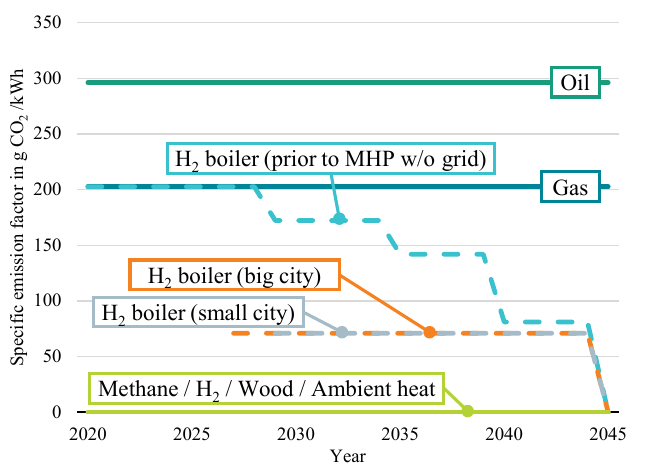}
    \caption{Development of infrastructure-independent specific GHG emission factors expressed in $\mathrm{g\,CO_2}$/kWh final energy for decentralised heating technologies and fuels. Besides conventional energy carriers (oil and natural gas), the figure includes renewable fuels and the regulatory emission accounting pathways for hydrogen boilers under different municipal heat planning and hydrogen infrastructure scenarios.}
    \label{fig:co2_emissions_decentral}
\end{figure}

\begin{figure}[pos=htbp]
    \centering
    \includegraphics[width=1\linewidth]{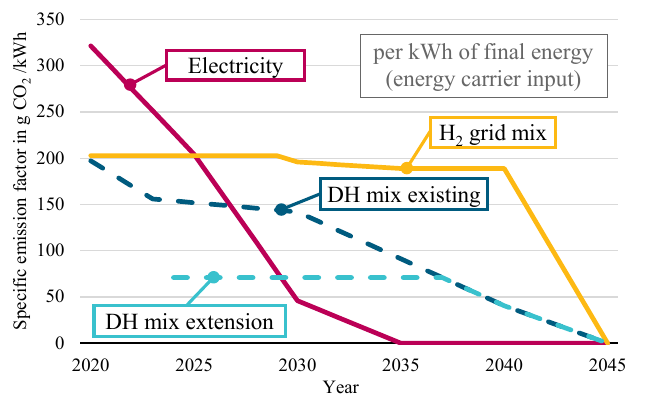}
    \caption{Development of infrastructure-dependent specific GHG emission factors expressed in $\mathrm{g\,CO_2}$/kWh final energy for electricity, district heating, and hydrogen grid supply. Emission factors reflect projected decarbonisation pathways of the respective energy infrastructures and distinguish between existing and newly expanded district heating networks.}
    \label{fig:co2_emissions_grids}
\end{figure}

\begin{figure}[pos=htbp]
    \centering
    \includegraphics[width=\linewidth]{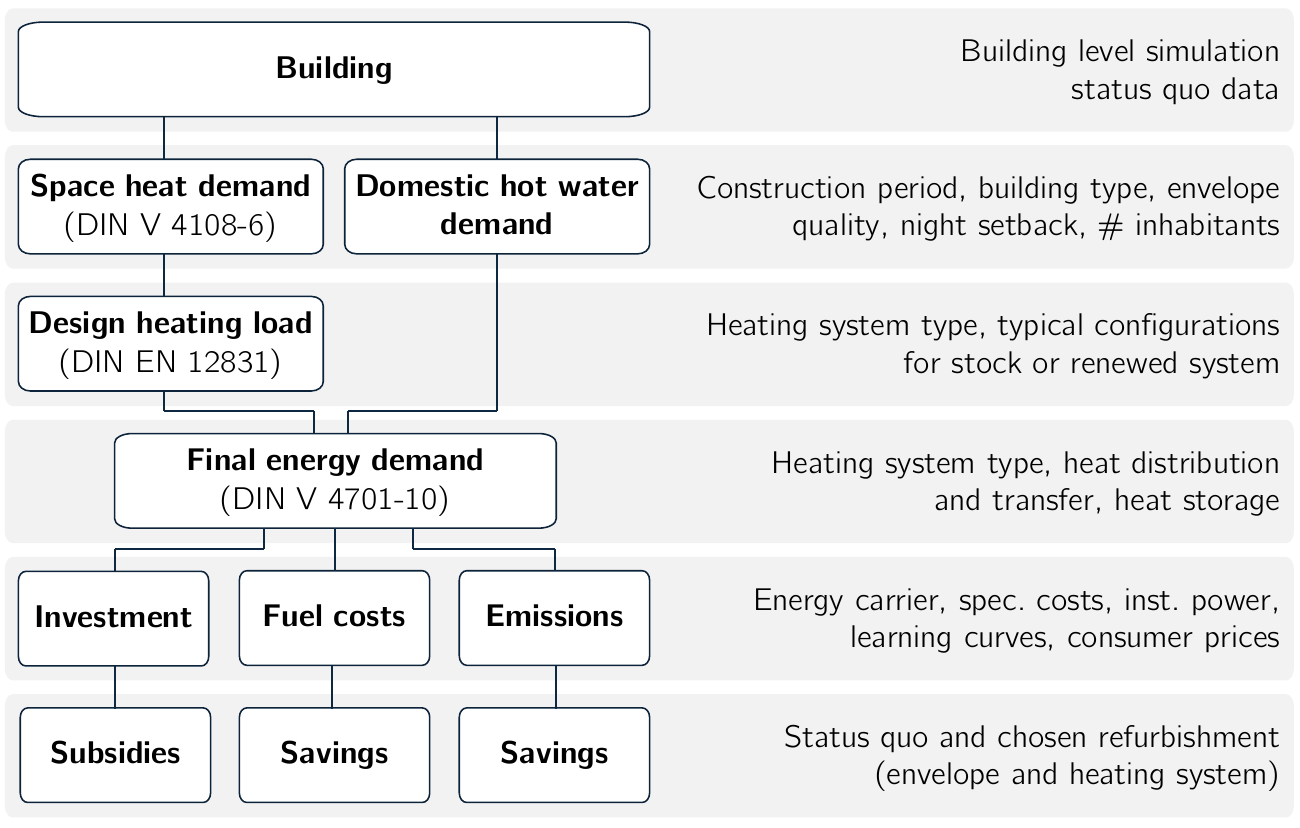}
    \caption{Attribute calculation for each building, for each refurbishment option, attributes in white boxes, the conditions accounted for on the light grey background. The calculations generally follow the applicable DIN standards, partly in simplified form. Starting from the building status quo and its fixed data (like the geometry, construction period, and building type), space heat demand is calculated according to relevant norms. Domestic hot water demand is calculated based on per-person assumptions and the number of inhabitants. Combined with the design heating load (necessary heating power to be installed), the final energy demand is derived. From these, installation costs, fuel consumption with expected fuel costs, and emissions are calculated, which eventually also lead to eligible subsidies and savings in costs and emissions.}
    \label{fig:attributes}
\end{figure}

The annual heating demand and final energy demand for the heating system -- comprising both space heating and domestic hot water -- are calculated in accordance with the German guidelines DIN\,V\,4108-6 and DIN\,V\,4701-10 (see Fig.~\ref{fig:attributes}).
Monthly solar radiation intensities and outdoor temperatures are incorporated into a component-level analysis using the building’s \textit{k}-values.
The annual primary energy demand is derived from the final energy demand by applying the primary energy factors specified in the GEG.
GHG emissions are computed from the final energy values, taking into account both the emission factors defined in the GEG and the evolution of infrastructure-dependent emission factors (e.g., for electricity, district heating, and gas mixtures), which are derived from macro-scale energy system planning models~\cite{Hartel.2022,Bottger.2021,Schmitz.2023}, statutory requirements, and complementary techno-economic assumptions (see Fig.~\ref{fig:co2_emissions_decentral} and Fig.~\ref{fig:co2_emissions_grids}).
Additionally, the required design load is calculated in accordance with DIN EN 12831, serving as the basis for cost calculations and as a key output parameter.

\subsubsection{Costs}
The model assesses investment and operating costs of refurbishment measures for both building envelope components and heating systems. It calculates investment expenditures as well as operating costs, including energy costs and $\mathrm{CO_2}$ taxes or certificate costs associated with the available options. Cost calculations account for general construction cost developments, represented by the construction price index and its projected evolution, as well as technology-specific learning effects that capture cost reductions associated with market diffusion, based on \cite{Energiewende.}.

The costs of building envelope refurbishment are derived from \cite{Hinz.10.08.2015}. They account for the initial building state and the resulting requirements for achieving a defined target performance level, including the necessary insulation thickness and associated area-dependent costs. Furthermore, a distinction is made between energy-related incremental costs and total investment costs.

Cost calculations for heating system installations are based on the construction cost database of the Construction Cost Information Centre of the German Chambers of Architects (BKI), from which technology-specific cost functions are derived. These functions express specific investment costs per installed kilowatt as a function of system size, thereby capturing economies of scale across different technologies.

For institutional owners, investment decisions are additionally evaluated using NPV calculations, with owner- and trigger-specific rates of return serving as discount rates and owner-role-specific evaluation horizons (see~\ref{app:npv}).

\subsubsection{Dynamics}
\label{sec:dynamics}
At the beginning of each time step, inflows and outflows from the building stock are modelled stochastically.
New buildings are added according to an assumed construction rate, while existing buildings are demolished according to an assumed demolition rate, with older and unrefurbished buildings having a higher probability of demolition.

For the rest of the building stock, investment decisions are triggered by discrete events such as changes in ownership or tenancy, the need for major maintenance, or the introduction of new regulatory requirements.
Continuous factors -- including fluctuations in cost trends, fuel prices, and policy updates -- dynamically modulate agents’ decision-making processes. 
The following subsections describe this dynamic decision process in detail.

\begin{figure}[pos=htbp]
\centering
\begin{tikzpicture}[
  scale=0.9,
  transform shape,
  font=\small,
  node distance=3.5mm and 10mm,
  >=Latex,
  block/.style={rectangle, draw, rounded corners=2pt, align=center,
                minimum width=40mm, minimum height=4.8mm},  
  decision/.style={diamond, draw, aspect=3.8, align=center,
                   inner sep=1pt, minimum width=40mm},
  note/.style={font=\footnotesize, align=center, text width=52mm}
]

\node[decision] (law) {Required by law?};
\node[decision, below=of law] (broken) {Part broken?};
\node[decision, below=of broken] (owner) {New owner?};
\node[decision, below=of owner] (tenant) {New tenant?};

\node[block, right=of law] (necess) {Refurbishment of broken/\\required part is necessary};
\node[block, right=of tenant] (no) {No refurbishment};
\node[block] at ($(necess)!0.5!(no)$) (all) {All refurbishment options\\are considered};

\draw[->] (law) -- node[right, font=\footnotesize]{No} (broken);
\draw[->] (broken) -- node[right, font=\footnotesize]{No} (owner);
\draw[->] (owner) -- node[right, font=\footnotesize]{No} (tenant);
\draw[->] (law) -- node[near start, above, font=\footnotesize]{Yes} (necess);
\draw[->] (broken.east) -- node[near start, above, xshift=-2mm, font=\footnotesize]{Yes} (necess.west);
\draw[->] (owner.east) -- node[near start, above, xshift=-2mm, font=\footnotesize]{Yes} (all.west);
\draw[->] (tenant.east) -- node[near start, above, xshift=-2mm, font=\footnotesize]{Yes} (all.west);
\draw[->] (tenant.east) -- node[near start, below, font=\footnotesize]{No} (no.west);

\end{tikzpicture}
    \caption{Trigger check and resulting refurbishment considerations. If there are legal obligations to replace a building part or if a part is broken, its refurbishment is necessary. More refurbishment options are considered additionally. If there is only a change in owner or tenant, refurbishment is completely optional.}
    \label{fig:flow_chart_triggers}
\end{figure}
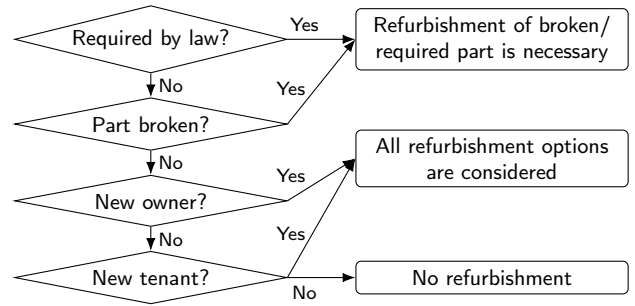

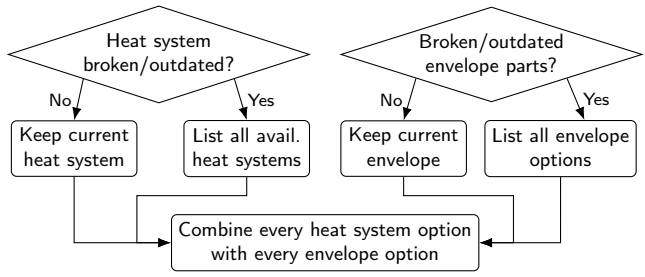
\begin{figure}[pos=htbp]
\centering
\begin{tikzpicture}[
  scale=0.83,
  transform shape,
  font=\small,
  node distance=3.5mm and 10mm,
  >=Latex,
  block/.style={rectangle, draw, rounded corners=2pt, align=center,
                minimum width=20mm, minimum height=4.8mm},
  decision/.style={diamond, draw, aspect=2.8, align=center,
                   inner sep=0pt, minimum width=48mm, minimum height=15mm},
  note/.style={font=\footnotesize, align=center, text width=52mm}
]

\node[decision] at (2.35, 3) (heatsys) {Heat system\\broken/outdated?};
\node[decision] at (7.65, 3) (env) {Broken/outdated\\envelope parts?};

\node[block] at (1.0, 1.5)  (keep_h) {Keep current\\heat system};
\node[block] at (3.75, 1.5) (list_h) {List all avail.\\heat systems};
\node[block] at (6.25, 1.5) (keep_e) {Keep current\\envelope};
\node[block] at (8.75, 1.5) (list_e) {List all envelope\\options};

\node[block]  at (5, 0) (combine) {Combine every heat system option\\with every envelope option};

\draw[->] (heatsys.south west) -- node[left, font=\footnotesize]{No} (keep_h.north);
\draw[->] (heatsys.south east) -- node[right, font=\footnotesize]{Yes} (list_h.north);
\draw[->] (env.south west) -- node[left, font=\footnotesize]{No} (keep_e.north);
\draw[->] (env.south east) -- node[right, font=\footnotesize]{Yes} (list_e.50);
\draw[->] (keep_h.south) -- (1, 0) -- (combine.west);
\draw[->] (list_h.south) -- (3.75, 0.75) -- (2.0, 0.75) -- (2, 0) -- (combine.west);
\draw[->] (keep_e.south) -- (6.25, 0.75) -- (8.0, 0.75) -- (8, 0) -- (combine.east);
\draw[->] (list_e.south) -- (8.75, 0) -- (combine.east);

\end{tikzpicture}
    \caption{General approach for the compilation of refurbishment options.}
    \label{fig:flow_chart_options}
\end{figure}

\paragraph{Refurbishment triggers}
At each five-year time step, the model evaluates for each agent whether critical events occur, see also Fig.~\ref{fig:fc_overview_tikz}. When an event is triggered (e.g., a change in tenancy or maintenance need; the need for major maintenance; or the introduction of new regulatory requirements), the corresponding decision-making process, either stochastic (for private individuals) or NPV-based (for institutional owners), is activated.
The trigger check is schematically depicted in Fig.~\ref{fig:flow_chart_triggers}.
Regulatory requirements as well as parts needing replacement at their end of life trigger mandatory refurbishments of the affected parts.
These triggers, as well as a change in owner or tenant, furthermore lead to the consideration of all available refurbishment options for the building envelope.
Legal obligations include mandatory roof insulation in case of a change of owner, according to current German law \cite{GEG}. In scenario variants, it is also possible to activate MEPS.
In the latter case, a specific energy demand exceeding MEPS requirements acts as a regulatory requirement.
\paragraph{Compilation of envelope and heating system refurbishment options}
The process of compiling the available options for heating system and building envelope is shown in Fig.\ \ref{fig:flow_chart_options}. The heating system is only replaced if it is necessary. For the building envelope, a pre-defined list of options is always considered, with mandatory inclusion of necessary replacements of parts that have reached their end of life or are required to be refurbished by existing regulations.
\paragraph{Envelope options}
For building envelope components, a predefined set of refurbishment options is considered.
A detailed overview of the considered refurbishment options and efficiency levels is provided in~\ref{app:env_opts}.
The entry \textit{old} indicates that the respective component remains unchanged and is not refurbished. The attainable refurbishment levels reflect different insulation standards: \textit{s1} corresponds to the minimum legal requirement for refurbishing the respective component, \textit{s2} denotes the minimum standard required to qualify for subsidies, and \textit{s3} represents the highest efficiency level below the passive house standard.
For components that require refurbishment, options that do not meet the required refurbishment level are excluded.
In addition, options are also excluded if the respective components do not satisfy a minimum component age.
Envelope refurbishment measures are only considered if the owners have sufficient financial resources.

\paragraph{Heating system options}
The available heating system options include air-source and ground-source heat pumps, with or without ventilation systems; pellet boilers, with or without solar thermal support; connection to district or local heating networks; and fossil fuel boilers (potentially $\mathrm{H_2}$-ready), with or without solar thermal support.
The availability of these options for a given building depends on a range of factors related to both the building itself and its surrounding environment.

The conditions determining the availability of heating system options are summarised in Table~\ref{tab:heat_opts}.
These include the availability of infrastructure at the building location, technical constraints of the building, the applicable regulatory framework, and additional model assumptions.

\begin{table*}[t]
    \caption{Factors influencing the availability of heating systems.
    Infrastructure availability and technical restrictions are
    building-specific and are assigned probabilistically based on
    Germany-wide statistical data, optionally supplemented by additional
    information or assumptions. Regulatory conditions depend on the
    progress of MHP, which is mandatory by mid-2026 for larger
    municipalities (more than 100,000 inhabitants) and by mid-2028 for
    smaller municipalities.}
    \label{tab:heat_opts}

    \centering
    \small
    \setlength{\tabcolsep}{5pt}

    \begin{tabularx}{\textwidth}{
        @{}
        >{\raggedright\arraybackslash}m{0.18\textwidth}
        >{\raggedright\arraybackslash}X
        @{}
    }
        \toprule

        Availability of infrastructure &
        \begin{tabitemize}
            \item District or local heating network
                  (potentially available only in a future year)
            \item Natural gas grid
                  (potentially available only until decommissioning or
                  conversion to hydrogen)
            \item Hydrogen grid
                  (potentially available only in a future year)
        \end{tabitemize}
        \\

        \midrule

        Technical restrictions &
        \begin{tabitemize}
            \item Ground-source heat pumps are only available if the
                  property area is sufficiently large
            \item Air-source heat pumps are only available if sufficient
                  distance to neighbouring buildings can be maintained
            \item Air-source heat pumps are only available for buildings
                  with up to eight flats
            \item Air-source heat pumps are only permitted if the building
                  envelope does not exceed a minimum efficiency threshold
                  defined for each building type
        \end{tabitemize}
        \\

        \midrule

        Regulatory conditions before MHP &
        \begin{tabitemize}
            \item Installation of new gas boilers is permitted, subject to
                  increasing shares of renewable fuels from 2029 onwards
            \item Replacement of oil boilers is permitted
        \end{tabitemize}
        \\

        \midrule

        Regulatory conditions after MHP &
        \begin{tabitemize}
            \item Immediate requirement of a 65\,\% renewable energy share
                  for all newly installed heating systems
            \item In general, only heat pumps, pellet boilers, or district
                  heating are permitted options (depending on availability)
            \item Exception: if gas-to-hydrogen conversion or expansion of
                  heating networks is planned, extended gas use and/or
                  $\mathrm{H_2}$-ready gas boilers may be permitted
        \end{tabitemize}
        \\

        \midrule

        Further assumptions &
        \begin{tabitemize}
            \item Connection to a heating network is not replaced by
                  another heating system
            \item Ground-source heat pumps are not replaced by other
                  heating systems
            \item Pellet boilers are available for all buildings, but the
                  total pellet budget is limited (first come, first served
                  in random scheduling order)
        \end{tabitemize}
        \\

        \bottomrule
    \end{tabularx}
\end{table*}

\paragraph{Final compatibility checks}
After the refurbishment options have been compiled, two final checks are performed, and options that do not satisfy these criteria are excluded.
First, refurbishment options exceeding the available budget are discarded.
Decisions are revised in random scheduling order, making the agent choose the next-best options without a pellet boiler.
Thus, the stochastic order affects the simulation result and is treated as a source of uncertainty (see Section~\ref{subsubsec:uncertainties}).
Second, air-source heat pumps are excluded for buildings whose specific heat demand, even after refurbishment, exceeds a building-type-specific threshold.
In hypothetical scenarios where MEPS for final energy demand are imposed, all options are additionally evaluated for compliance with the respective MEPS requirements.
\paragraph{Decision process}
The model features two decision-making modes.
For agents representing private building owners (owner-occupiers, private landlords, or homeowner associations), a random utility approach is applied.
Institutional owners (private property companies, public authorities, or cooperatives) are modelled to decide based on a NPV assessment of the different options.
The foundation for both rationales is the detailed calculation of the attributes of each refurbishment option, which is illustrated in Fig.~\ref{fig:attributes}. 

\textbf{Random utility valuation.}
Each refurbishment option is evaluated using the class-specific WTP values
for the attributes listed above. The owner's utility $U$ of option $A_n$ is then calculated as
\begin{align}
    U(A_n)=\sum_j \left( \mathrm{WTP}_j^{(q)}\ Y_j^{(n)}\right)\ +\epsilon_n
\end{align}
where $j$ enumerates all attributes, $q$ is the latent class of the given owner, $\mathrm{WTP}_j^{(q)}$ is the WTP value for attribute $j$, depending on the latent class $q$, $Y_j^{(n)}$ is the attribute value of attribute $j$ in option $n$, and $\epsilon_n$ is a realisation of an extreme value distributed random variable, accounting for model limitations.
After the utility has been calculated for all options, the option with maximal utility
\begin{align}
    A^{\mathrm{opt}}=\operatorname{argmax}_n U(A_n)
\end{align}
is selected.

\textbf{Net present value.}
For institutional owners, investment decisions are based on a NPV assessment, in which future costs and revenues are discounted over the evaluation period. Depending on the ownership role, different cash flows are considered (Table~\ref{tab:owner_decision_logic}), while the corresponding NPV assumptions and parameter values are summarised in Table~\ref{tab:npv_assumptions}. All institutional owners account for investment costs and available subsidies.

For self-use-oriented assessments, the economic benefit results from discounted savings in future operating costs over the service life of the respective building component, including avoided fuel expenditures and applicable $\mathrm{CO_2}$ costs.

For rental-oriented assessments, future rental income is evaluated under two rental situations. During an ongoing tenancy, landlords are assumed to increase rents annually towards the local reference rent, irrespective of refurbishment measures. Consequently, these regular rent adjustments do not constitute an economic benefit of energy-related refurbishment. Instead, the additional revenue attributable to refurbishment is limited to the difference between the rent increase permitted through the modernisation levy and the rent increase that could have been achieved through the regular adjustment to the local reference rent. To reflect that the legally permitted modernisation levy is typically shared between energy-related and other modernisation measures, the model assumes that only 50\,\% of the admissible levy is allocated to energy-related refurbishment investments.

Following a tenant change, the legal restrictions governing rent increases during an ongoing tenancy no longer apply. Instead of assuming unrestricted rent setting, the model adopts a conservative modelling assumption: rents are increased to the local reference rent and a limited modernisation levy is additionally considered. This assumption is intended to represent the economic value of refurbishment while avoiding an overestimation of rental revenues and reflecting that a refurbished dwelling would otherwise not be marketable. Furthermore, landlords account in both cases for the owner share of future $\mathrm{CO_2}$ costs. An overview of the decision approaches and the corresponding evaluation criteria for the different owner types is provided in Table~\ref{tab:owner_decision_logic}.

\begin{table*}[t]
\centering
\caption{Decision approaches and evaluation criteria for different owner types.}
\label{tab:owner_decision_logic}
\small
\renewcommand{\arraystretch}{1.15}

\begin{tabularx}{\textwidth}{
    l
    >{\raggedright\arraybackslash}p{3.0cm}
    >{\raggedright\arraybackslash}p{3.2cm}
    >{\raggedright\arraybackslash}X
}
\toprule
\textbf{Owner group} &
\textbf{Owner type} &
\textbf{Decision approach} &
\textbf{Evaluation criteria} \\
\midrule

\multirow{3}{*}{Private owners}

& Owner-occupiers
& WTP-based (self-use)
& Fuel costs (including CO$_2$ costs) \\ \\

& Homeowner associations
& WTP-based (self-use)
& Fuel costs (including CO$_2$ costs) \\ \\

& Private landlords
& WTP-based (rental)
& Owner share of CO$_2$ costs; modernisation levy \\

\midrule

\multirow{3}{*}{Institutional owners}

& Private property companies
& NPV-based (rental-oriented)
& Investment costs net of subsidies; owner share of CO$_2$ costs;
reference-rent adjustment; modernisation levy \\

& Public authorities
& NPV-based (rental- or self-use-oriented)
& Rental-oriented: investment costs net of subsidies, owner share of
CO$_2$ costs, reference-rent adjustment, and modernisation levy;
self-use-oriented: investment costs net of subsidies and avoided fuel costs (including CO$_2$ costs) \\

& Cooperatives
& NPV-based (self-use-oriented)
& Investment costs net of subsidies; avoided fuel costs (including CO$_2$ costs) \\

\bottomrule
\end{tabularx}
\end{table*}

\paragraph{Cancellation}
An owner-type-specific cancellation rate is applied to both envelope and heating system refurbishments following a refurbishment trigger event. The rate determines whether the initiated refurbishment process proceeds to the evaluation of refurbishment options and captures exogenous factors not explicitly modelled, such as shortages of installers or other qualified workers, material constraints, uncertainty regarding future regulatory developments, or additional financial burdens on the owners.
A cancelled refurbishment of a building component that has reached the end of its service life can be interpreted as a minimal repair or patching measure.
Refurbishment cancellation is excluded under the following conditions:
\begin{itemize}
    \item If a gas grid is decommissioned and the building relies on gas heating, replacement of the heating system cannot be cancelled.
    \item If the refurbishment includes the installation of an air-source heat pump and this installation is not cancelled, the corresponding envelope refurbishment cannot be cancelled, in order to prevent the use of air-source heat pumps in buildings with excessively high energy demand.
\end{itemize}
\subsection{Input and output}
\subsubsection{Data sources}\label{subsubsec:data_sources}
The building stock, including ownership structures and socio-demographic characteristics, is derived from SUF data of the Federal Statistical Office (Statistisches Bundesamt).
Details such as the initial refurbishment state of the building envelope and technical systems are inferred from construction year and component lifespans.
Key parameters include:
\begin{itemize}
    \item probability distributions for random utility in private decision-making
    \item cost trends for maintenance and renovation \citep{koch_2021}
    \item energy source price trajectories
    \item regulatory intervention schedules (e.g., timing and scope of the gas grid phase-out)
    \item availability of heat sources and grid infrastructure
\end{itemize}

A summary of model data differentiation and corresponding sources is given in Table \ref{tab:data_sources}.

\subsubsection{Data processing }
\label{sec:data_proc}
Missing input data for landlord characteristics and building data, such as the exact number of flats, construction year, or exact household income, are supplemented stochastically.
Since there is no information on whether houses with one or two flats are terraced or detached, this information must also be assigned probabilistically.
The stochastic process ensures the model reflects uncertainty in the input data and real-world variability.
For example, while the SUF database provides detailed information on the status of current heating systems, it lacks direct data on ownership for rental properties.
In such cases, ownership is probabilistically assigned based on income data and rental revenue indicators.
Furthermore, the heating system information is processed and consolidated to remove implausible combinations from the empirical questionnaire data set.

Special attention is paid to the building parts' ages and corresponding thermal insulation quality, as this critically influences the energy demand.
To impute a building part's age, the year of construction and the part's typical lifetime \cite{bte} are combined, assuming that the part needs replacement every time it reaches its end of life. 
This time is component-specific. For example, the average service life of a roof in Germany is approximately 80 years, whereas windows are on average replaced after around 40 years. Based on the year of construction and the expected lifetime of each component, a (probable) last replacement year is derived  using Weibull-distributed service lives for the respective building components. The insulation quality is in turn determined from what was typical for this replacement year (depending on past regulation and building standards) as reported in \cite{iwu:2016}.

\subsubsection{Key assumptions}
\label{sec:key_assumptions}
The main assumption for the development of privately owned buildings is the extrapolation of the empirically derived WTP values into the future, i.e.\ the assumption that the WTP values do not change over time.
Moreover, we normally assume that agents act myopically in their valuation of operating costs, in particular regarding $\mathrm{CO_2}$ pricing.

With regard to residential buildings, the model relies on archetype buildings we derived from Germany-wide LoD2 data, differentiated by building type and number of dwelling units.
This representation is, however, only an approximation.
In cases where all buildings within a specific region are modelled explicitly using local data (see Section~\ref{sec:building_stock}), the building characteristics are instead directly obtained from the corresponding local datasets.

We also assume a constant rate in new construction as well as demolition, which can be varied as an input parameter and is typically set to averages of historic data.
The share of highly efficient buildings among all new buildings is estimated at one third and is approximated by the historic ratio of efficiency-subsidised new buildings to total new construction.

Another fundamental assumption is the envelope quality estimation based on a synthetic building history as described in Section~\ref{sec:data_proc}.
\subsubsection{Scenario parameters}
The main differences between the modelled scenarios concern the regulatory framework. By default, the model reflects the legislation and funding conditions in force as of May 2026, including \cite{BEG, GEG, WPG}.
Additionally, further policy measures can be activated to evaluate their effects (see Section~\ref{sec:environment} for details).

\subsubsection{Output parameters}
The simulation records these metrics at the building level:
\begin{itemize}
    \item timing and outcome of renovation and heating system replacement decisions
    \item final energy and heat demand
    \item the impact of these decisions on CO\textsubscript{2} emissions
    \item investment costs and changes in monthly fuel expenses
    \item funding and apportioned costs for landlords
\end{itemize}
Post-processing routines exist to aggregate these data per building type, building age class, ownership type, income group, installed technology, and energy source as appropriate.
At the system level, consumption of energy sources such as biomethane and biomass, as well as spent funding, are recorded.

\subsection{Quality assurance}

\subsubsection{Calibration}
The model is calibrated to reproduce realistic refurbishment dynamics and decision patterns.
Calibration is primarily achieved by adjusting the assumed service lives of heating systems and building components, as well as by introducing minimum component ages before replacement or refurbishment becomes feasible.
This prevents an overestimation of refurbishment frequencies, as observed service lives in practice typically exceed technical lifetime assumptions. 

In addition, refurbishment cancellation rates are calibrated and differentiated by owner type.
In particular, higher cancellation rates are assigned to private owners, reflecting financial constraints and implementation effort, and to homeowner associations, capturing inertia in collective decision-making processes.
These parameters are calibrated against historical developments and subsequently extrapolated into future scenarios.

A further calibration parameter is the decay parameter, which defines the share of the building stock assigned to declining housing market conditions. It serves as a proxy for whether a local housing market is characterised by a tenant or landlord market. This allows the model to account for the fact that energy refurbishment measures may not be economically viable in weak housing markets or may require additional policy support. The classification is based on observed rent levels and associated probabilities and is calibrated to reproduce historically observed development patterns.

\subsubsection{Validation}
Model outcomes are validated by cross-referencing historical heating-system replacement rates and energy consumption data. Representative survey data on the German residential building stock are used as a reference for the observed age distribution of heating systems and for estimating plausible replacement ages and replacement rates \cite{BDEW.2023.WieHeiztDeutschland}. These are complemented by technology sales and market-development data to validate the historical diffusion and replacement of heating technologies \cite{BDH.2025.Marktentwicklung}. Building-level energy consumption is used to calibrate the model, while state-level final energy consumption data are applied to validate energy use by energy carrier. Quantitative measures (e.g., correlation coefficients) are employed to assess whether the simulation results adequately reproduce observed trends.

\subsubsection{Limitations }
In addition to the key assumptions described in Section~\ref{sec:key_assumptions}, the model relies on several simplifying assumptions and is subject to data-related limitations.
Building-specific characteristics, such as the refurbishment state, ownership structure, heat supply type and age, and the age of individual building components, are not fully observable at the individual building level and must therefore be estimated.
The available data basis in Germany is limited, particularly with respect to spatial resolution and differentiation by building age class and building type.
As a result, key attributes are derived from aggregated statistics and assigned to individual buildings using stochastic methods.
This introduces uncertainty in the representation of building-specific conditions, including the identification of private landlords, as well as in the modelling of decision-relevant characteristics.
These limitations, combined with aggregation effects and incomplete information in the underlying datasets, constrain the accuracy of the building-level representation. At the same time, aggregated statistics such as heat supply types are then missing for out-of-sample validation.

Additional uncertainty arises from the calibrated model parameters.
Key parameters, such as service lives, minimum component ages, and refurbishment cancellation rates, are calibrated based on historical observations and extrapolated into future scenarios, which may not adequately capture future behavioural changes or structural shifts.
Moreover, the representation of market conditions, including the distinction between tenant and landlord markets, relies on probabilistic classifications, introducing further uncertainty in modelling market dynamics and the feasibility of refurbishment measures.

Several input parameters, including future energy prices, regulatory developments, and infrastructure availability, are treated as exogenous and associated with considerable uncertainty, particularly in long-term scenario projections.
This also applies to willingness-to-pay parameters, which represent a snapshot in time and may change in response to societal and political developments, such as energy price shocks, regulatory uncertainty, or macroeconomic disruptions.
Furthermore, agent-to-agent interactions are currently not represented. Consequently, social influence or peer effects do not endogenously affect refurbishment decisions or the willingness-to-pay parameters of individual agents. Incorporating such interaction effects remains subject to future work.

\subsubsection{Uncertainty} \label{subsubsec:uncertainties}
Several parts of the model are based on Monte Carlo methods.
Random sampling is used for two main purposes. First, it is applied to generate building- and agent-specific attributes from different data sources that cannot be directly linked at the individual building level.
Second, it is used to represent decisions or events that are outside the explicit scope of the model, such as the passing of building owners.
A detailed analysis of these uncertainties is beyond the scope of the present study and will be addressed in subsequent work.

\subsubsection{Sensitivity}
Sensitivity analyses are performed by the one-factor-at-a-time (OFAT) method for fuel prices, cost trends, regulatory stringency and building sample sizes to assess their influence on the transition dynamics. Implausibly sensitive parameters were not detected. Given the computational cost of global sensitivity analysis sampling the model output over a wide range of parameter values, we currently restrict sensitivity studies to OFAT. Global sensitivity analyses are left to future work.

\subsection{Additional information }
\subsubsection{Implementation }
The framework is implemented in Python using the Mesa framework \cite{mesa.2020}.
Input data management is based on Structured Query Language (SQL) using PostgreSQL and Excel spreadsheets, calculations mainly use \verb|pandas| \cite{pandas.2020} and \verb|numpy| \cite{numpy.2020}, and visualisation of results applies \verb|matplotlib| \cite{matplotlib.2007} and \verb|seaborn| \cite{seaborn.2021}.
\subsubsection{Access}
We are currently in the process of transforming the model into open-source software.
Information on how to access code and documentation will be published on the project website (\href{https://www.iee.fraunhofer.de/de/anwendungsfelder/energiewirtschaft/quartier-stadt-region/AgentHomeID.html}{https://www.iee.fraunhofer.de/AgentHomeID}) as soon as available. Furthermore, this website contains details for contacting current developers.
Access to data from \cite{SUF} for research purposes requires accreditation.
\subsubsection{Funding and contributors}
Funding providers are listed below in the acknowledgements and contributions in the CRediT statement.
\subsubsection{Areas of application}
\begin{figure*}[pos=htbp]
    \centering
    \includegraphics[width=1\linewidth]{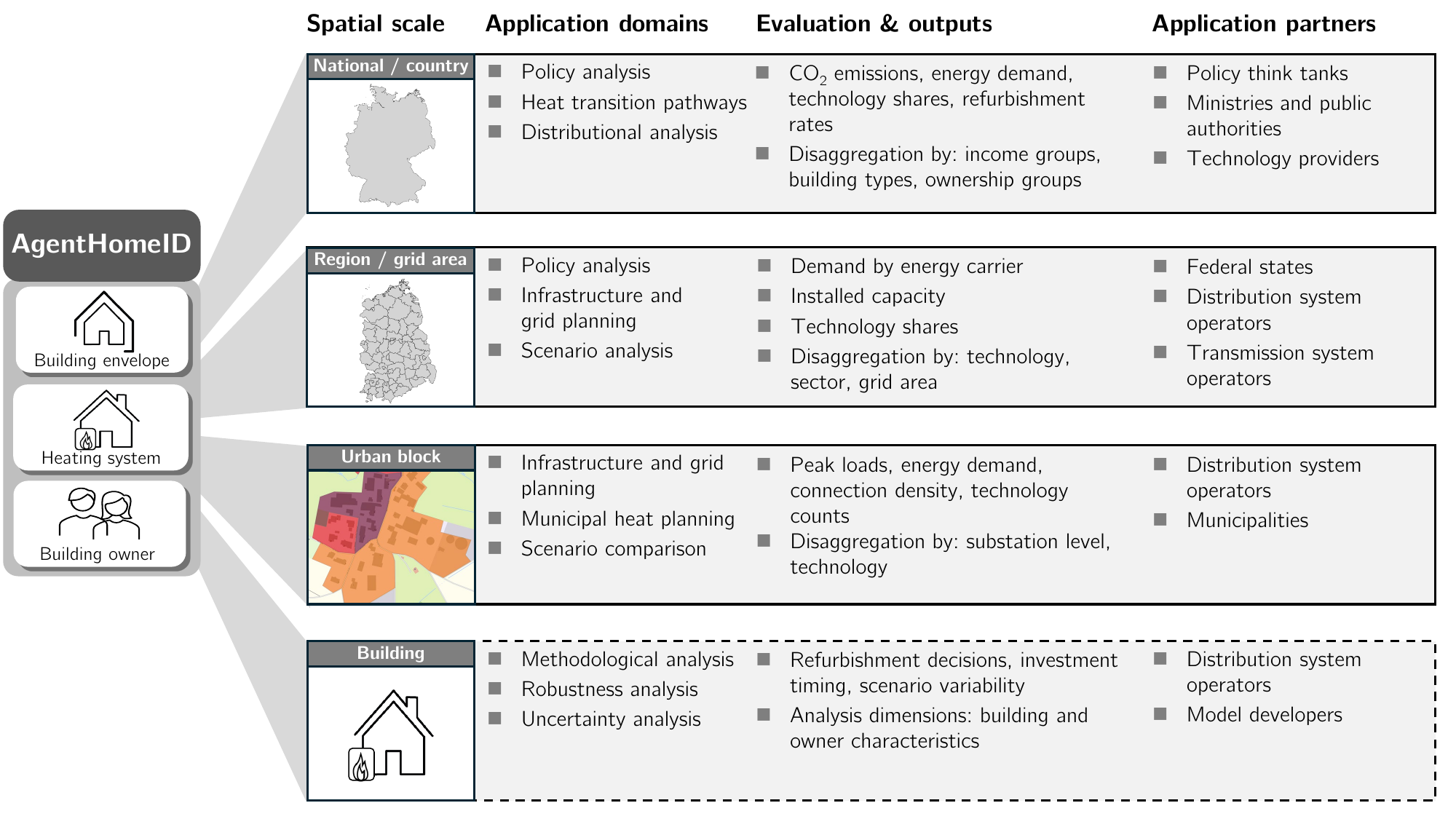}
    \caption{Overview of the model’s application domains across different spatial scales, ranging from national and regional levels to distribution grid areas and local substation regions. Depending on the application and research question, the model provides varying levels of granularity and supports different types of analysis.}
    \label{fig:application_areas}
\end{figure*}

The model can be applied across multiple spatial scales and analytical contexts, ranging from individual buildings and urban blocks to entire grid regions, states, or national building stocks, as shown in Fig.~\ref{fig:application_areas}.
It enables the evaluation of policy measures, the analysis of long-term heat transition pathways, and the assessment of distributional and burden-sharing effects across different income groups, building types, and ownership structures.
Furthermore, the model supports infrastructure and grid planning by providing detailed outputs on energy demand, peak loads, and technology deployment.
Its flexible structure also allows for robustness analyses and theoretical investigations of policy designs under varying assumptions.
\section{Model applications and results}
\label{sec:applications}

This section presents the model's main application domains across different
spatial scales.
For each domain, we outline the typical research questions, the relevant application partners, the suitable evaluation metrics, and the required level of analytical granularity.
Depending on the application context, the model supports different forms of evaluation, ranging from aggregated system-level assessments to highly disaggregated analyses at the level of buildings, owners, and occupants.

The following subsections present each application domain by spatial scale.
For each, we report representative results that demonstrate the corresponding model capabilities.
Further methodological detail and extended analyses for the individual case studies are available in the corresponding project-specific publications.

\subsection{National-scale applications}
At the national scale, the model is used to analyse long-term transformation pathways of the building stock, assess the effects of policy measures, and evaluate aggregate indicators such as energy demand, $\mathrm{CO_2}$ emissions, installed technologies, and refurbishment activities. In addition to these aggregated assessments, the model allows for disaggregated analyses by building type, construction period, and ownership group. Owing to the explicit representation of owners and occupants at the building level, national-scale applications also enable the analysis of distributional effects and burden sharing across owner and tenant groups and, at a more detailed level, across income groups. Typical application partners at this scale include ministries, federal authorities, policy-oriented think tanks, and technology providers.
\subsubsection{Use case: Socially just heat transition in Germany}
\paragraph{Application context and motivation}
This use case addresses the transformation of the German residential heating sector under alternative policy and infrastructure pathways, with a particular focus on the interaction between decarbonisation targets and distributional effects. The transition is shaped by a combination of regulatory requirements, subsidy schemes, carbon pricing, and infrastructure developments, all of which influence investment decisions of heterogeneous building owners. At the same time, concerns about affordability and social equity have become increasingly important in the policy debate. Assessing both the technical transformation of the building stock and the distribution of costs and benefits across socio-economic groups therefore requires an integrated modelling approach that captures behavioural decision-making and heterogeneous agent characteristics. The analysis builds on two complementary applications of the AgentHomeID model, focusing on regulatory design and infrastructure-driven transformation pathways \cite{Ganal.2025, Ganal.2026}.
\paragraph{Research question}
The overarching research question is how different policy and infrastructure designs affect the decarbonisation trajectory of the residential building stock and the distribution of economic impacts across owner and income groups. Specifically, the analysis investigates (i) the role of regulatory requirements for heating system replacement and (ii) the impact of infrastructure interventions, such as gas grid decommissioning, on technology adoption, energy demand, and cost distribution.
\paragraph{Scenario design}
The analysis combines two complementary scenario perspectives applied to the German residential building stock over the period up to 2045. First, alternative regulatory designs for heating systems are compared, including scenarios with and without binding renewable energy requirements for newly installed systems. Second, infrastructure-driven transformation pathways are analysed by contrasting a business-as-usual (BAU) scenario with an ambitious climate protection (ACP) scenario that includes a regulated phase-out of gas distribution networks. Both scenario perspectives are embedded in a consistent framework of socio-economic assumptions, including carbon pricing trajectories, subsidy schemes, and infrastructure availability. This combined design allows the effects of specific policy instruments to be isolated while capturing their interaction with behavioural decision-making. Furthermore, the impact of a mandated renewable share for new heating systems is investigated as a sensitivity analysis. For further details, see \cite{Ganal.2025, Ganal.2026}.
\paragraph{Evaluation approach}
The evaluation integrates system-level and distributional perspectives. Key technical indicators include final and useful heat demand, technology shares, heating system replacement dynamics, and the composition of energy carriers over time. In addition, distributional impacts are assessed by analysing cost burdens, subsidy allocation, and investment activity across different owner types (e.g. owner-occupiers, private landlords, institutional owners) and income groups.

Results are evaluated at multiple levels of aggregation, ranging from the entire building stock to disaggregated analyses by energy carrier, building characteristics, ownership type, and income group. Furthermore, combined dimensions, such as the interaction between ownership structure and income distribution, are considered to capture distributional effects in greater detail.

This range of evaluation perspectives is particularly well suited to the agent-based modelling approach applied in this study. The explicit representation of individual buildings and heterogeneous agents enables the consistent aggregation of results across different dimensions while preserving behavioural and structural heterogeneity. As a result, the model links system-level transformation dynamics with micro-level decision processes, thereby providing insights into both overall transition pathways and their distributional implications across socio-economic groups.
\paragraph{Key findings}
\textit{Technical system impacts:}
The results show that both regulatory stringency and infrastructure interventions have a strong influence on the pace and direction of the heating transition. Removing binding renewable heating requirements leads to a substantially higher final energy demand, driven by the continued adoption of fossil-based systems with lower upfront investment costs.

\begin{figure}[pos=htbp]
    \centering
    \includegraphics[width=\linewidth]{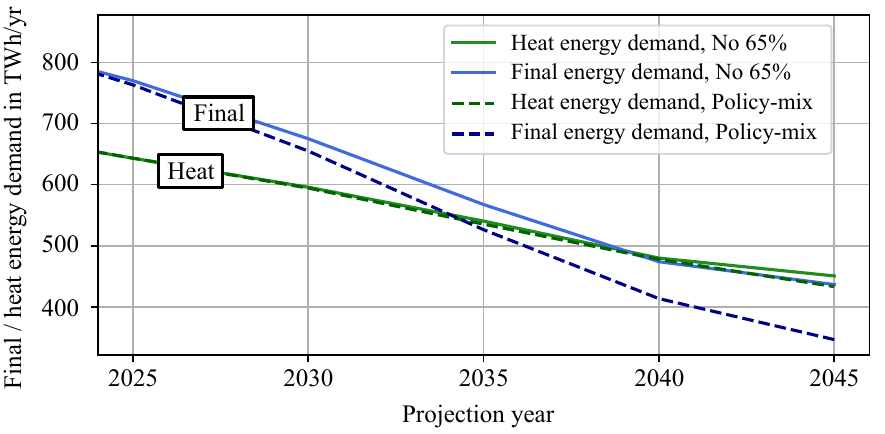}
    \caption{Comparison of final energy demand and useful heat demand under a policy-mix scenario and a scenario without binding renewable heating requirements.}
    \label{fig:energy_demand_policy}
\end{figure}

This effect is illustrated in Fig.~\ref{fig:energy_demand_policy}, which compares the development of final and useful heat demand under a policy-mix scenario with binding renewable requirements and a scenario without such constraints.
While useful heat demand evolves similarly across scenarios, reflecting comparable levels of building envelope refurbishment, final energy demand diverges substantially.
In the absence of regulatory constraints, the increased reliance on less efficient fossil-based heating technologies leads to significantly higher energy consumption over time.
This highlights the importance of technology choice, rather than solely building efficiency improvements, for achieving overall energy demand reductions \cite{Ganal.2026}.

In contrast, scenarios with regulated gas grid decommissioning accelerate the phase-out of fossil fuels and substantially increase the uptake of heat pumps.

\begin{figure}[pos=htbp]
    \centering
    \includegraphics[width=\linewidth]{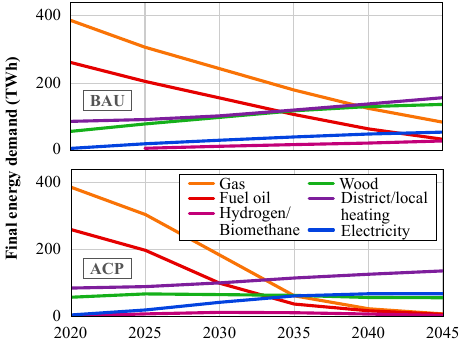}
    \caption{Development of final energy demand by energy carrier for BAU and ACP scenarios \cite{Ganal.2025}.}
    \label{fig:energy_demand}
\end{figure}

The evolution of final energy demand by energy carrier is illustrated in Fig.~\ref{fig:energy_demand}.
The comparison between the BAU and the ACP scenario shows a persistent reliance on fossil fuels in the BAU case, with gas and oil still contributing significantly to total energy demand in 2045.
In contrast, the ACP scenario achieves a complete phase-out of fossil fuels, accompanied by a strong increase in electricity and district heating demand as key pillars of decarbonisation \cite{Ganal.2025}.

\begin{figure}[pos=htbp]
    \centering
    \includegraphics[width=\linewidth]{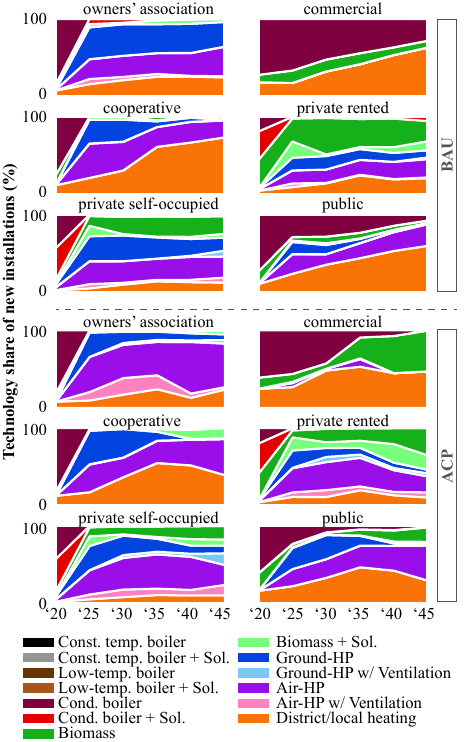}
    \caption{Technology shares of newly installed heating systems over time for different ownership groups in BAU and ACP scenarios \cite{Ganal.2025}.}
    \label{fig:heating_tech}
\end{figure}

These differences are primarily driven by changes in heating technology adoption. As shown in Fig.~\ref{fig:heating_tech}, the ACP scenario leads to a substantially higher penetration of heat pumps across all ownership groups, while fossil-based systems remain prevalent in the BAU scenario.
The transition dynamics differ across owner types: owner-occupiers tend to adopt higher-efficiency technologies earlier, whereas private landlords and commercial actors show a stronger preference for lower upfront investments.
The combination of regulatory constraints and infrastructure changes in the ACP scenario effectively shifts these preferences towards low-carbon technologies, resulting in a faster and more comprehensive transformation of the heating system stock \cite{Ganal.2025}.

This divergence across owner types emerges directly from the empirically estimated, class-specific WTP: it is a behavioural pattern that ownership-undifferentiated models cannot represent.

\begin{figure}[pos=htbp]
    \centering
    \includegraphics[width=\linewidth]{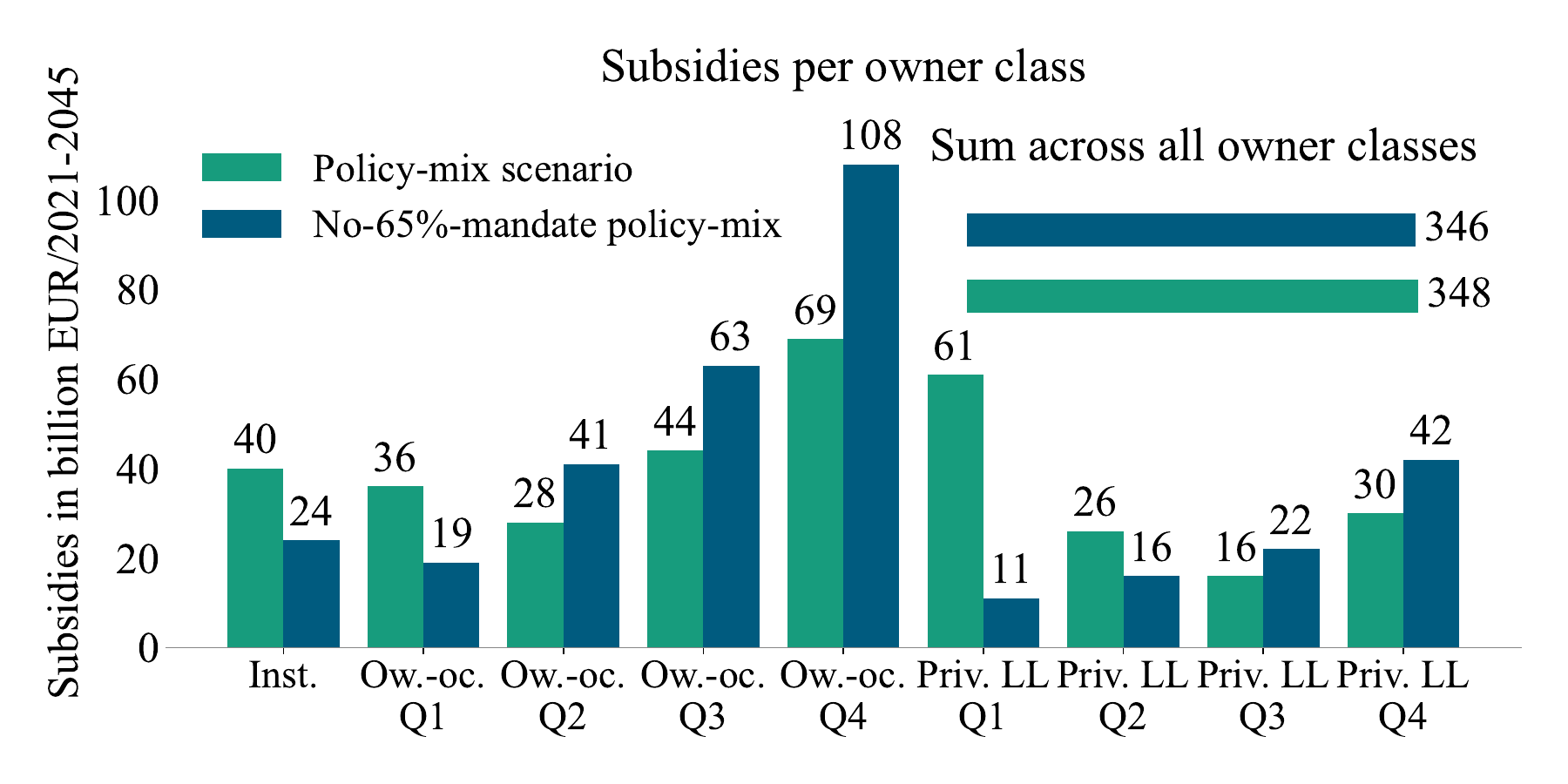}
\caption{Cumulative subsidy volumes received by different owner groups and income quartiles between 2021 and 2045 under the policy-mix scenario and the policy-mix scenario without the 65\,\% renewable mandate. The figure illustrates how subsidy allocation shifts across owner categories when fossil heating technologies remain available. Q1--Q4 denote income quartiles from lowest (Q1) to highest (Q4)
\cite{Ganal.2026}.}
    \label{fig:subsidy_distribution}
\end{figure}

\textit{Distributional impacts:}
Beyond the technical system transformation, the analysis reveals pronounced distributional effects across socio-economic groups. The interaction of regulatory design and behavioural decision-making leads to differentiated investment patterns and cost burdens across income groups and ownership structures. The distribution of cumulative subsidy volumes across owner groups and income quartiles is shown in Fig.~\ref{fig:subsidy_distribution}

\begin{figure}[pos=htbp]
    \centering
    \includegraphics[width=\linewidth]{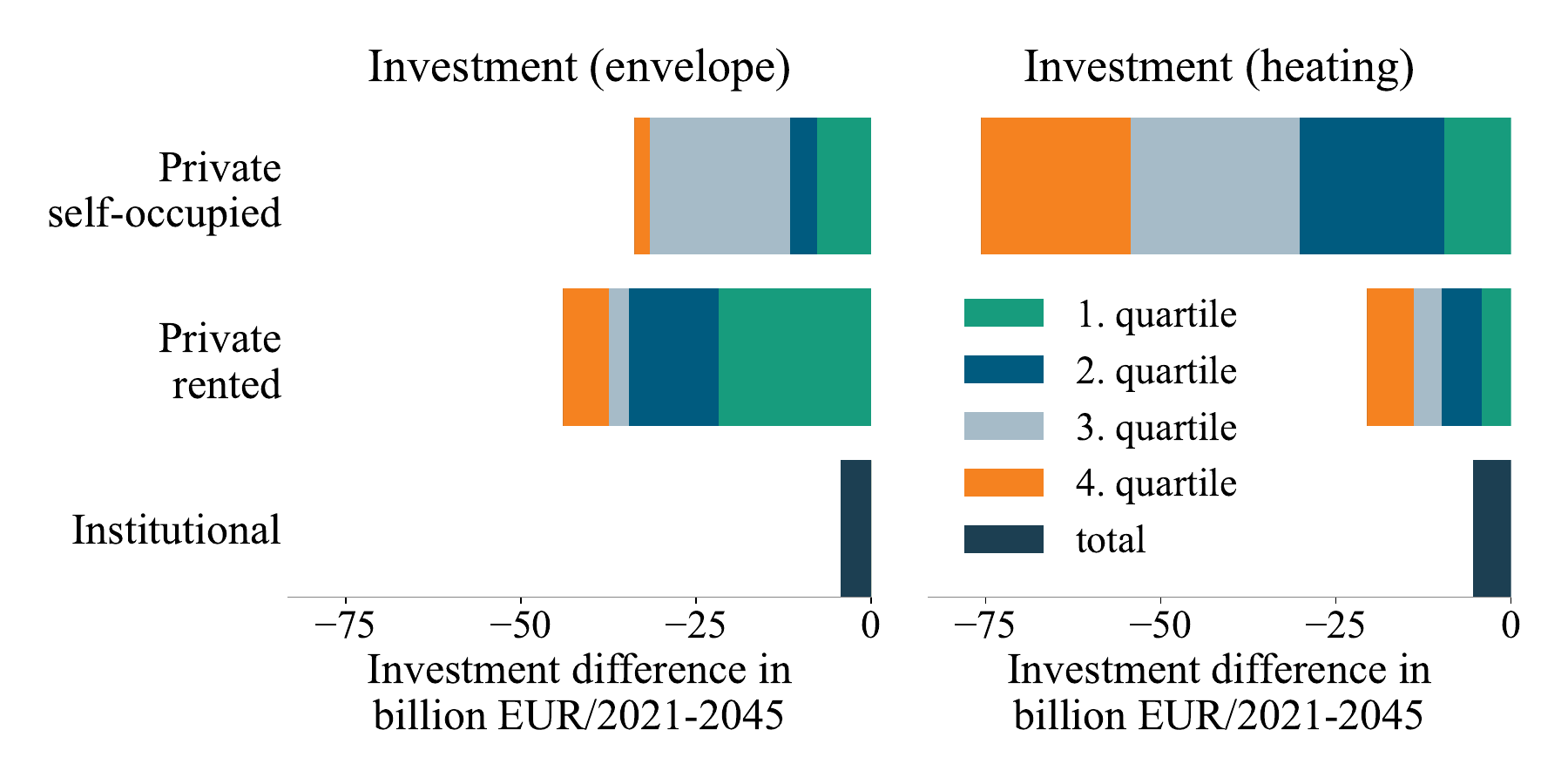}
    \caption{Investment shortfalls in building envelope and heating systems by income group and ownership type \cite{Wauer.2025c}.}
    \label{fig:investment_gap}
\end{figure}

These patterns are further reflected in the distribution of investment shortfalls across income groups, as shown in Fig.~\ref{fig:investment_gap}. The figure highlights that lower-income households exhibit substantially lower investment levels in both building envelope measures and heating systems.
In particular, the lowest income quartiles show substantial underinvestment, especially among private landlords and self-users. This indicates that financial constraints and behavioural factors limit participation in the transition, leading to unequal access to efficiency improvements and long-term cost savings \cite{Wauer.2025c}.

Overall, the results demonstrate that policy design not only determines aggregate decarbonisation outcomes but also strongly influences the distribution of costs, benefits, and investment opportunities across society. While infrastructure-driven interventions such as gas grid decommissioning can reduce long-term system costs and stabilise energy demand, distributional challenges remain and require targeted policy measures to ensure a socially balanced transition.
\paragraph{Implications}
The combined results highlight that policy design and infrastructure planning jointly determine both the effectiveness and the social acceptability of the heat transition.
Regulatory frameworks that allow continued investment in fossil-based technologies may reduce short-term investment pressure but lead to higher long-term system costs and adverse distributional outcomes.
In contrast, coordinated policy and infrastructure strategies, including clear regulatory signals and managed infrastructure phase-outs, can accelerate decarbonisation while improving cost stability and equity.
The use case demonstrates the capability of the \textsc{AgentHomeID} model to simultaneously analyse technological transformation pathways and distributional impacts, thereby supporting the design of policy mixes that balance climate targets with social considerations.
\paragraph{Further details}
Detailed descriptions of the scenario assumptions, model implementation, and quantitative results are provided in the corresponding publications \cite{Ganal.2025, Wauer.2025, Wauer.2025b, Wauer.2025c, Ganal.2026}.
\subsection{Regional and grid-level applications}
At the regional and grid level, the model is applied to develop medium- to long-term scenarios for geographically defined areas, such as federal states as well as transmission or distribution grid regions. At the level of federal states, the model supports the analysis and monitoring of region-specific transformation targets under alternative policy scenarios. At the grid level, it enables the derivation of key indicators relevant for infrastructure development, such as energy demand by energy carrier, installed capacities, and technology shares. These outputs provide an important basis for planning under both realistic and target-driven assumptions. Typical application partners at this scale include federal states, distribution system operators, and transmission system operators. Both types of applications, at the regional and grid level, have already been implemented in real-world contexts within the \textsc{AgentHomeID} framework. 
\subsubsection{Use case: Regional energy transition scenarios for distribution grid planning (Planning Region East, Germany)}
\label{sec:prost}
\paragraph{Application context and motivation}
The transformation of the energy system, particularly the electrification of the heating and transport sectors, poses significant challenges for distribution system operators (DSOs). In Germany, DSOs are required to develop regional scenarios as a basis for long-term grid expansion planning (according to Section 14d of the German Energy Industry Act
(Energiewirtschaftsgesetz, EnWG)). These scenarios must provide consistent projections of future grid-connected assets while ensuring alignment with national climate targets and regulatory frameworks.

In this context, the presented use case focuses on the development of regionalised energy transition scenarios for the Eastern planning region of Germany. The overall modelling approach follows a modular framework in which different technologies and sectors are represented by dedicated models. 

The agent-based model presented in this paper is specifically applied to simulate the development of decentralised heating systems and the resulting heat demand of buildings, while simultaneously providing the basis for identifying potential district heating areas. In particular, it is used to model the uptake of decentralised heat pumps and the evolution of building-level heat demand, which are key drivers of future electricity demand in the distribution grid.

The results of this model are integrated with other sector-specific modelling approaches (e.g.\ for renewable generation, electric mobility, or industrial demand) to derive a consistent regional scenario that reflects both national targets and regional specificities \cite{horst_2025_ut, Fischer.2025, Regionalszenario2023, Regionalszenario2025}.
\paragraph{Research question}
The use case addresses the following key questions:
(i) How will electricity demand and production evolve at the regional level under the given scenario assumptions?
(ii) What is the expected contribution of decentralised heat pumps, district
heating supplied by heat pumps, and building heat demand to future electricity
consumption?
(iii) How do infrastructure developments, such as district heating expansion or hydrogen network availability, influence the electrification of the heating sector?
(iv) Which implications arise for distribution grid expansion and long-term investment planning?
\paragraph{Scenario design}
The scenario design follows a hybrid approach combining national-level targets with region-specific adjustments.
Long-term developments are anchored in national targets (e.g. network development plan (NEP) scenario B for 2045), while intermediate years (2030 and 2035) are derived based on expected technology-specific development trajectories and regional boundary conditions.

The scenario integrates assumptions on technology diffusion, infrastructure development, and regulatory frameworks.
Within this setup, the agent-based model contributes the bottom-up simulation of building-level refurbishment and heating system decisions, determining the development of decentralised heat pumps, heat network connections and the associated heat demand.
By design, this use case develops a single planning-relevant reference scenario rather than a set of contrasting pathways, since its purpose is to provide a consistent basis for infrastructure decisions rather than to explore scenario sensitivity.
The combination of exogenous national assumptions with endogenous building-level modelling ensures consistency across sectors while capturing regional heterogeneity \cite{Fischer.2025, Regionalszenario2025}.
\paragraph{Evaluation approach}
The evaluation focuses on deriving regionally aggregated indicators relevant for infrastructure planning. Key outputs include electricity demand, installed capacities of key technologies (e.g.\ heat pumps), and the spatial distribution of energy demand across administrative units (e.g.\ Nomenclature of Territorial Units for Statistics level 3, NUTS-3) and grid regions.

In contrast to more fine-grained urban applications, the analysis at this level prioritises consistency with and embedding in national scenario frameworks, as well as the robustness and interpretability of aggregated indicators, over highly detailed local resolution.
Results are therefore primarily evaluated at regional and sub-regional scales, while still being grounded in building-level simulations.

The modelling approach enables linking bottom-up decision processes with top-down scenario constraints, providing a consistent framework for translating national energy transition pathways into regionally differentiated demand projections.
This allows DSOs to assess infrastructure needs under different scenario assumptions while maintaining comparability with overarching planning frameworks.
At the same time, the approach facilitates the comparison and evaluation of top-down scenario inputs provided by transmission system operators (TSOs), thereby supporting consistency checks and analytical alignment between transmission-level and distribution-level planning.
\paragraph{Key findings}
The results highlight a substantial increase in electricity demand across the region, primarily driven by the electrification of the heating sector. In particular, the uptake of decentralised heat pumps represents a key driver of future load growth, with pronounced regional differences reflecting variations in building structure, existing heating systems, and local boundary conditions.

\begin{figure}[pos=htbp]
    \centering
    \includegraphics[width=1\linewidth]{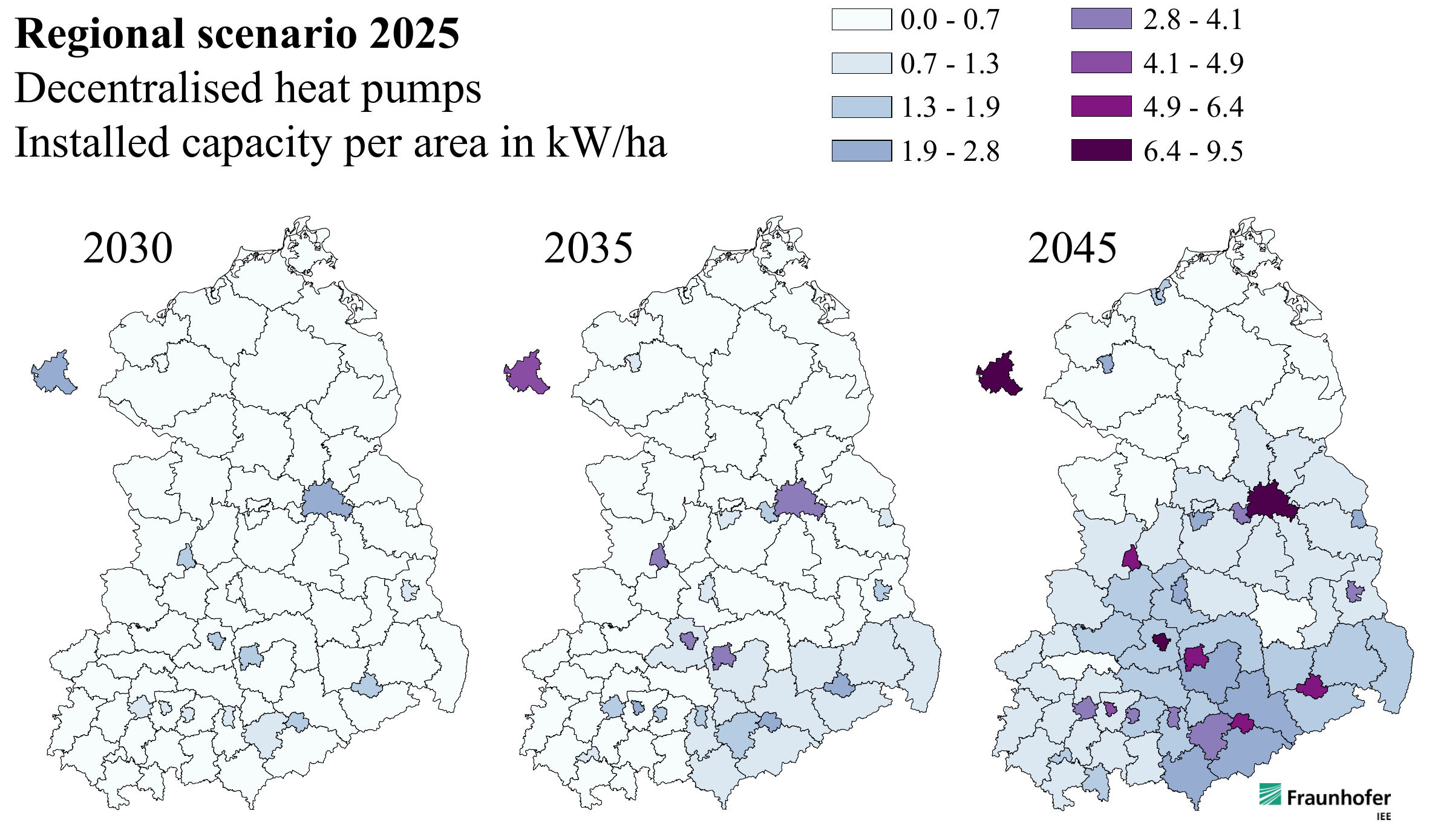}
    \caption{Projection of installed electrical capacity of decentralised heat pumps (air-source and ground-source heat pumps) in the Eastern planning region of Germany at NUTS-3 level for the years 2030, 2035, and 2045 (excluding electric backup heaters) \cite{Regionalszenario2025}.}
    \label{fig:hp_regional}
\end{figure}

The spatial and temporal development of decentralised heat pump capacities is illustrated in Fig.~\ref{fig:hp_regional}. The results show a strong increase in installed electrical capacity of air-source and ground-source heat pumps between 2030 and 2045, with a clear concentration in regions characterised by favourable building structures and limited district heating availability. The agent-based simulation reveals that this development is not spatially uniform, but instead follows region-specific patterns driven by heterogeneous investment decisions and local infrastructure conditions. These results underscore the importance of capturing bottom-up dynamics to identify regional hotspots of electrification and corresponding grid expansion needs.

\begin{figure}[pos=htbp]
    \centering
    \includegraphics[width=1\linewidth]{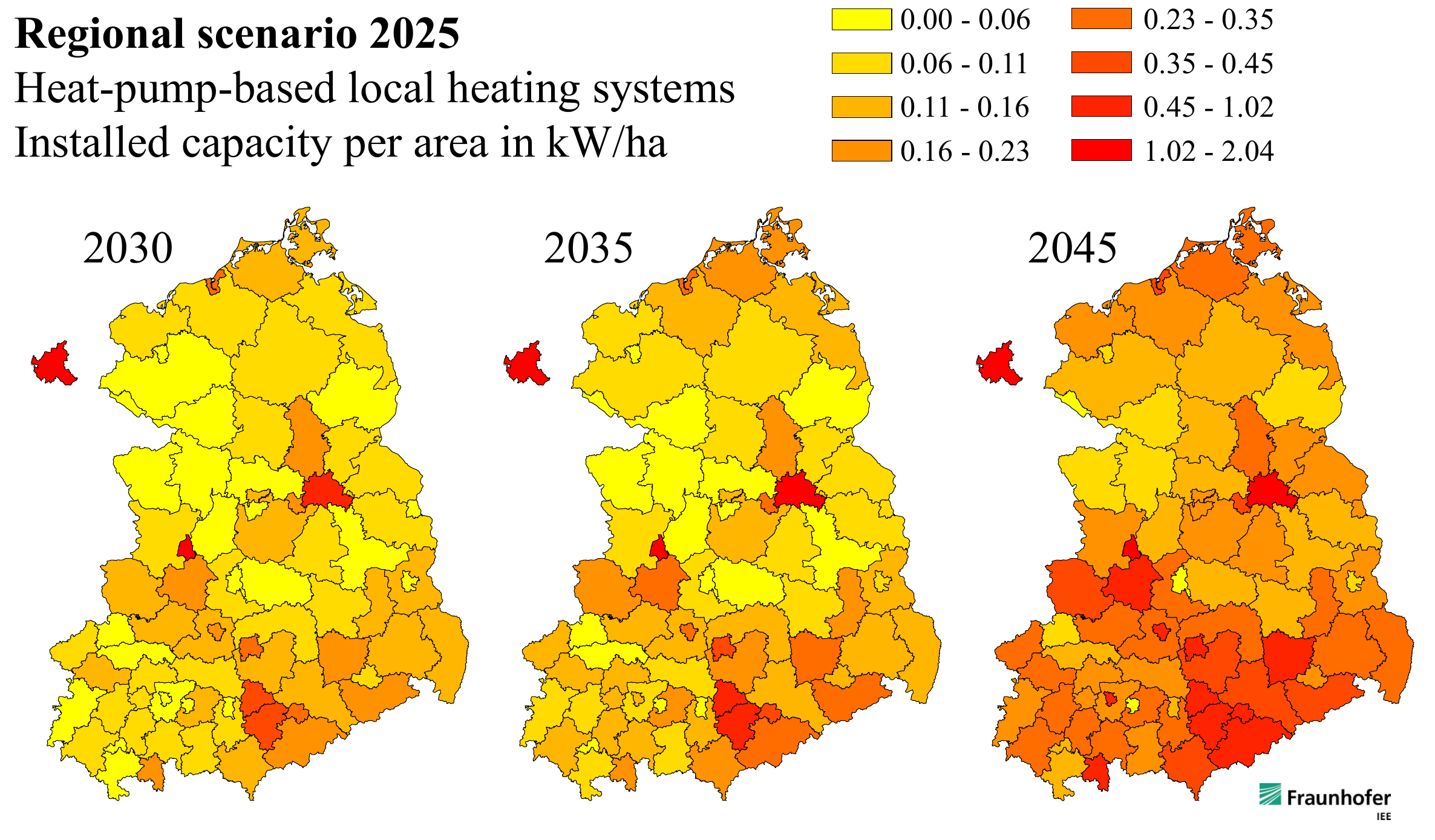}
    \caption{Projection of installed electrical capacity of local heating systems in the Eastern planning region of Germany at NUTS-3 level for the years 2030, 2035, and 2045 (excluding combined heat and power units) \cite{Regionalszenario2025}.}
    \label{fig:dh_capacity}
\end{figure}

In parallel to decentralised electrification, the analysis also shows a substantial expansion of district and local heating systems. As illustrated in Fig.~\ref{fig:dh_capacity}, the installed electrical capacity of local heating networks increases steadily over time, particularly in regions with higher heat demand densities.
This reflects the model-based identification of economically viable areas for network expansion, where building density and demand clustering favour centralised heat supply solutions.
The interaction between decentralised heat pump adoption and district heating expansion plays a crucial role in shaping regional electricity demand trajectories.

\begin{figure}[pos=htbp]
    \centering
    \includegraphics[width=0.5\linewidth]{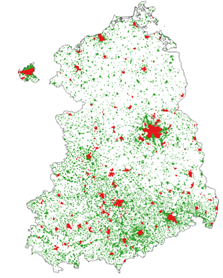}
    \caption{Model-based projection of the spatial distribution of district heating (red) and local heating (green) supply areas in 2045 \cite{Regionalszenario2025}.}
    \label{fig:dh_areas}
\end{figure}

The spatial allocation of district and local heating supply areas in 2045 is shown in Fig.~\ref{fig:dh_areas}. The results indicate that district heating (red) is predominantly concentrated in urban centres, while local heating systems (green) extend into peri-urban and selected rural areas. This differentiation reflects the underlying economic and spatial constraints of network expansion and highlights the importance of combining centralised and decentralised solutions in regional heat transition strategies.

Beyond the spatially resolved results, aggregated indicators derived from the agent-based simulation are compared with existing regional scenarios and current national scenario frameworks (e.g. NEP scenarios). In particular, total installed heat pump capacity and the number of systems are benchmarked against top-down projections.

The comparison demonstrates how building-level modelling influences both the overall technology diffusion and its spatial distribution compared with aggregated scenario approaches. The observed deviations reflect the influence of behavioural decision-making, heterogeneous building characteristics, and local constraints that are not captured in purely aggregated approaches. The explicit representation of individual buildings and owner-specific decision-making constitutes a key distinction of AgentHomeID compared with aggregated building stock models (Tables~\ref{tab:model_comparison_a} and ~\ref{tab:model_comparison_b}).

These differences are particularly relevant for grid planning, as they affect both total load levels and the spatial concentration of demand.
\paragraph{Implications}
The findings demonstrate the importance of combining national scenario frameworks with region-specific, bottom-up modelling approaches for infrastructure planning.
While national scenarios provide essential boundary conditions, their translation into regional demand projections requires a detailed representation of local building structures, infrastructure availability, and behavioural decision-making.

The results highlight that aggregated top-down projections may not fully capture regional dynamics, particularly with respect to the spatial distribution and timing of electrification.
By incorporating building-level decision processes, the modelling approach enables a more differentiated assessment of demand development, including the identification of regional hotspots of grid expansion needs.

For distribution system operators and planning authorities, the approach provides a robust basis for deriving planning-relevant reference scenarios, assessing future grid expansion requirements, and ensuring consistency with national planning frameworks. In addition, the results support the comparison and evaluation of transmission-level scenario assumptions, thereby contributing to improved alignment between transmission and distribution system planning.

Overall, the combined modelling framework supports the development of consistent, data-driven regional scenarios that account for both structural and behavioural drivers of the energy transition and align short-term planning needs with long-term decarbonisation targets.
\paragraph{Further details}
A detailed description of the methodology, scenario assumptions, and full results can be found in the corresponding project report and related publications \cite{horst_2025_ut, Fischer.2025, Regionalszenario2023, Regionalszenario2025}.
\subsection{Urban block-level applications}
At the urban block level, the model enables detailed analyses of localised interactions between buildings, infrastructure, and stakeholders. Based on these simulations, key indicators such as electrical peak loads, heat grid connection densities, energy demands, and their temporal development can be derived at the level of local substations or higher-voltage substations. These outputs support grid operators in planning and expanding network infrastructure in the presence of multiple decentralised consumers and generators. In addition, the model supports municipal heat planning by identifying priority areas for district heating or hydrogen networks and by deriving the indicators required to assess the transformation of the heat sector at block level over time. This includes the comparison of various scenarios and the derivation of actionable recommendations. Typical application partners at this scale include municipalities, distribution grid operators, and municipal utilities.
\subsubsection{Use case: High-resolution grid load assessment and municipal heat planning at urban block level}
\paragraph{Application context and motivation}
At the urban block level, planning requirements shift from aggregated scenario analysis towards spatially and temporally detailed assessments of infrastructure needs. Distribution system operators require high-resolution information on load development at the level of local substations and higher voltage interfaces in order to identify concrete grid expansion requirements \cite{Regionalszenario2023, Regionalszenario2025}. At the same time, German municipalities are required by law to conduct municipal heat planning, which necessitates a detailed spatial representation of heat demand, technology potentials, and future infrastructure development at the level of individual blocks or districts \cite{horst_2025_ut, Geiger.04.02.2026}.

The presented use case addresses both application contexts by extending the regional modelling results to a high spatial resolution. On the one hand, it enables detailed load assessments for electricity grids based on decentralised electrification. On the other hand, it supports municipal heat planning by providing spatially explicit information on technology deployment, infrastructure potentials, and transformation pathways.
\paragraph{Research question}
The use case addresses two complementary sets of questions:
(i) How do decentralised heating technologies, in particular heat pumps, translate into electrical loads at the level of local substations and higher voltage interfaces, and what are the resulting peak load requirements for grid expansion?
(ii) How can building-level simulation results be used to identify suitable areas for district and local heating systems and to support the development of municipal heat plans at block level?
\paragraph{Scenario design}
The use case builds on the regional reference scenario described in Section~\ref{sec:prost} and applies it at a higher spatial resolution. No alternative scenario pathways are considered. Instead, the focus lies on the spatial and temporal disaggregation of a consistent scenario to derive planning-relevant indicators at local level.

Building-level simulation results from the agent-based model are used as the basis for both applications. These results include the development of heating technologies (e.g. air-source and ground-source heat pumps), refurbishment activities, and the evolution of heat demand. In addition, the modelling of district and local heating networks is refined iteratively based on spatial demand density and infrastructure feasibility, allowing the identification of potential expansion and densification areas.
\paragraph{Evaluation approach}
The evaluation focuses on translating building-level simulation results into spatially and temporally resolved indicators for infrastructure planning. For the application to electricity grids, installed electrical capacities of decentralised heat pumps are aggregated at the level of local substations and higher voltage interfaces. Technology-specific load profiles are subsequently scaled to the building-specific electricity demand calculated by \textsc{AgentHomeID}. Simultaneity effects are considered as a function of the number of heat pumps connected to the respective grid node, allowing probabilistic peak loads for small heat-pump collectives and increasingly smoothed load profiles for larger aggregates to be represented \cite{Krause.2026}.

The resulting time series are used to derive maximum load conditions at different grid levels. Importantly, the relevant grid-planning quantity is therefore not simply the sum of installed heat-pump capacities or individual technology peaks, but the coincident electrical load resulting from the temporal operation of the connected systems. In combination with additional technologies such as photovoltaic systems, battery storage, and electric mobility, integrated load profiles can subsequently be constructed to identify maximum load situations arising from the interaction of multiple technologies. This enables the spatial identification of grid expansion requirements across different voltage levels.

\begin{figure}[pos=htbp]
    \centering
    \includegraphics[width=\linewidth]{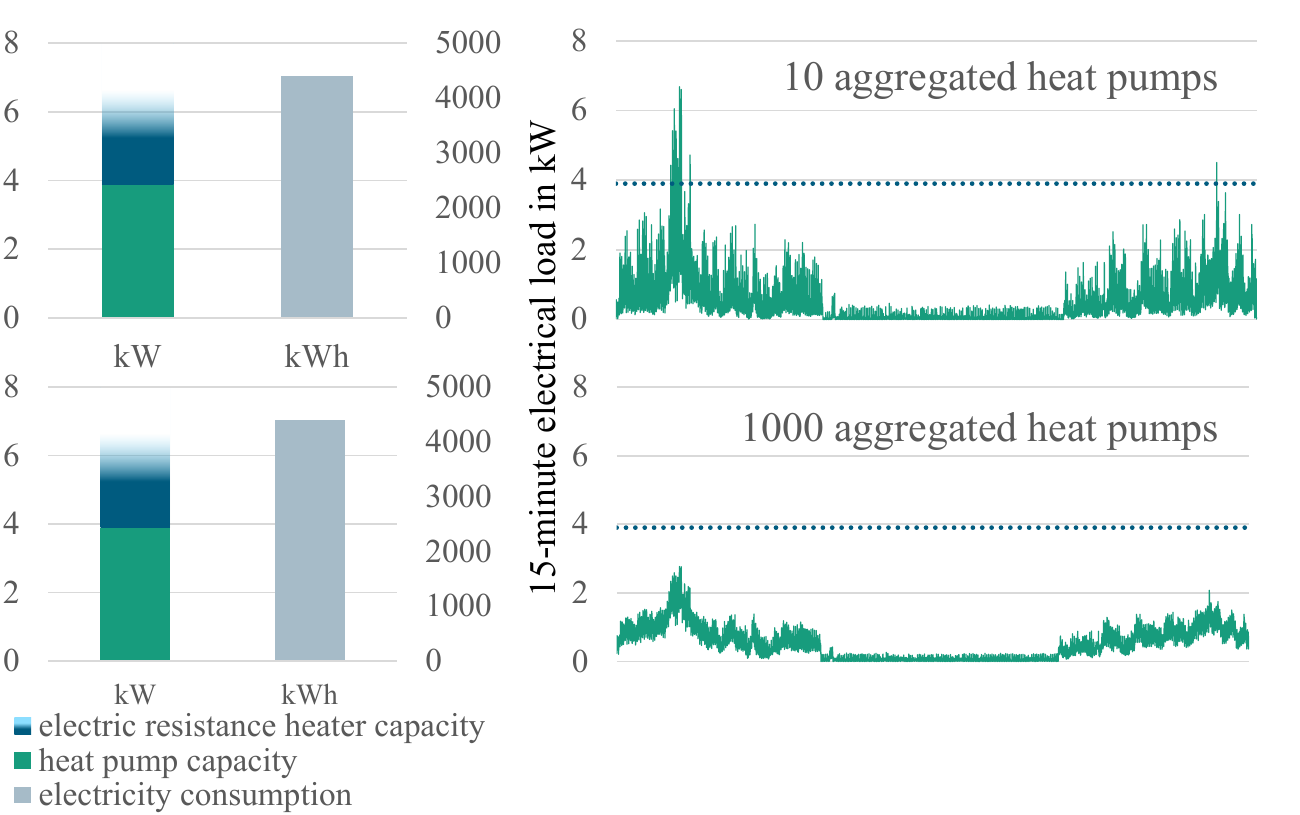}
    \caption{Illustration of the derivation of air-source heat-pump loads for grid-planning applications in 2045. The left panels show installed electrical heat-pump capacity, auxiliary electric resistance-heater capacity, and annual electricity consumption. The right panels show the resulting annual load profiles relative to the installed heat-pump capacity for aggregates of 10 (top) and 1000 (bottom) air-source heat pumps. Small aggregates exhibit substantially higher probabilistic peak loads, whereas increasing aggregation results in pronounced simultaneity effects and smoother load profiles.}
    \label{fig:hp_load_profiles}
\end{figure}

For municipal heat planning, the evaluation focuses on spatial indicators at urban block level. These include technology-specific potentials derived from technical boundary conditions and infrastructure modelling, as well as scenario-based developments of heating systems and building refurbishment. The results are provided in a spatially explicit format and can be explored via an interactive web-based viewer, allowing users to analyse technology deployment, heat demand, and infrastructure development over time \cite{Geiger.04.02.2026}.\footnote{\url{https://maps.iee.fraunhofer.de/kwp/}}
\paragraph{Key findings}
The results show that the spatial disaggregation of building-level simulations leads to a highly heterogeneous distribution of electrical loads at the level of local substations. While some substations experience only moderate increases in load, others exhibit strong growth driven by clustered heat pump adoption and high local heat demand. This highlights the importance of high-resolution modelling for identifying local grid bottlenecks that are not visible in aggregated analyses.

The generation of time-resolved load profiles further demonstrates the importance of simultaneity for translating installed heat-pump capacities into grid-relevant peak loads. As illustrated in Fig.~\ref{fig:hp_load_profiles}, small heat-pump collectives can exhibit pronounced probabilistic load peaks, including contributions from auxiliary electric resistance heaters, whereas aggregation across a larger number of systems substantially smooths the resulting load profile. Consequently, installed electrical capacity alone is not sufficient to determine the maximum load relevant for grid planning. The approach allows the building-level technology development simulated by \textsc{AgentHomeID} to be translated into time-resolved load assumptions for individual grid nodes and subsequently into grid planning scenarios.

When heat-pump loads are combined with other decentralised technologies such as electric vehicles, photovoltaic systems, and battery storage, the relevant system peak additionally depends on the temporal coincidence of the individual load and generation profiles. Maximum system load therefore does not necessarily coincide with the individual technology peaks or their arithmetic sum. This technology-spanning simultaneity is particularly relevant for assessing future grid reinforcement requirements.

In the context of municipal heat planning, the results demonstrate that the combination of building-level simulation and infrastructure modelling allows for a detailed identification of suitable areas for district and local heating systems. High-density urban areas are primarily allocated to district heating, while local heating solutions emerge in peri-urban and selected rural areas. At the same time, the simulation of owner decisions provides insights into the expected development of decentralised technologies and refurbishment activities, enabling a more realistic assessment of future heat demand reductions and technology uptake.

\begin{figure}[pos=htbp]
    \centering
    \includegraphics[width=\linewidth]{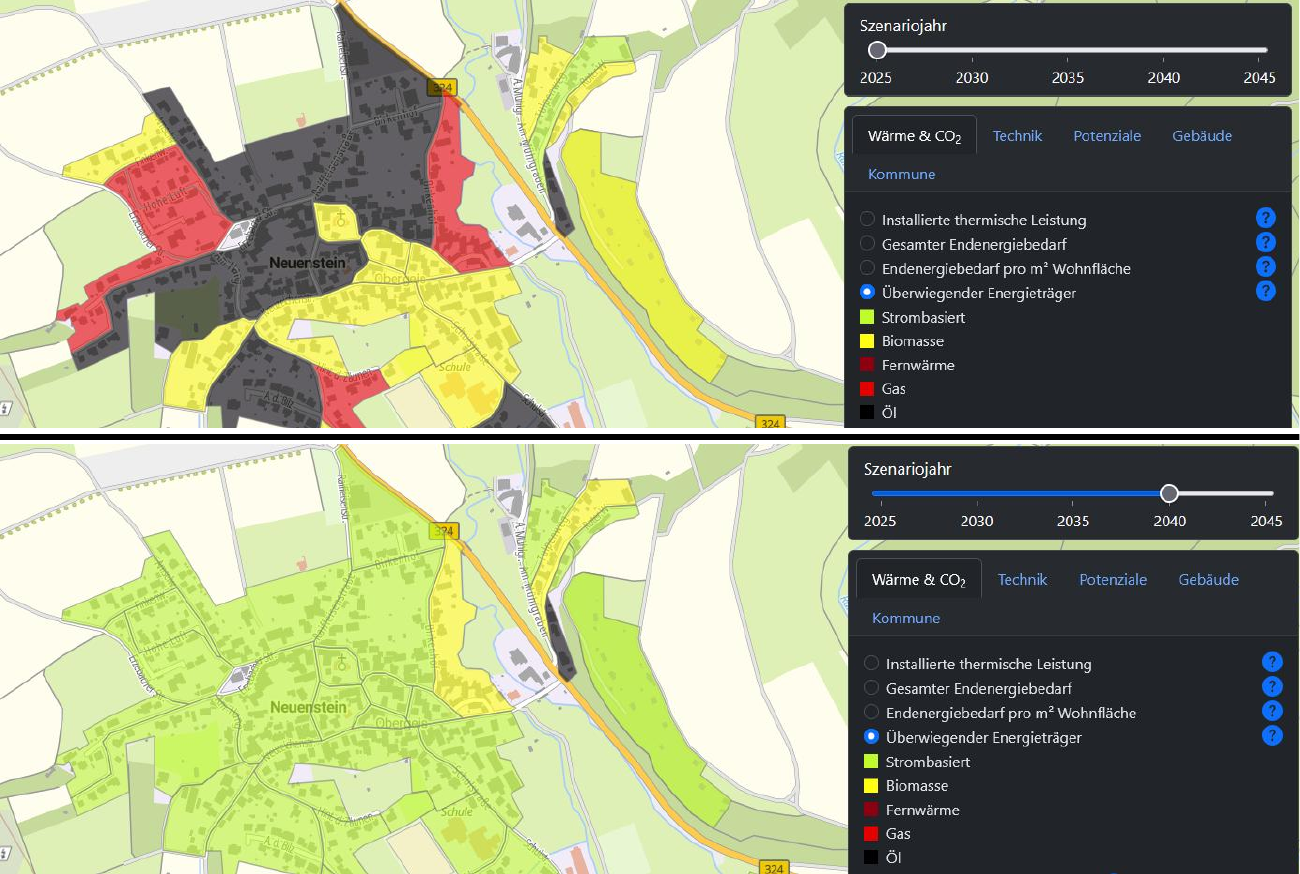}
    \caption{Example of the web-based planning tool showing the dominant energy carrier at urban block level for the base year 2025 (top) and the projected target year 2040 (bottom) \cite{Geiger.04.02.2026}.}
    \label{fig:kwp_viewer}
\end{figure}

An example of the spatial resolution and temporal development of the results is illustrated in Fig.~\ref{fig:kwp_viewer}, which shows the dominant energy carrier at urban block level for the base year 2025 and the projected target year 2040 as visualised in the web-based planning tool \cite{Geiger.04.02.2026}. The comparison illustrates both the strong spatial heterogeneity of the existing heat supply structure and its transformation over time. This fine-grained representation supports municipal heat planning by enabling transformation needs and potential priority areas for infrastructure development to be identified at urban block level.

\paragraph{Implications}
The findings demonstrate that high-resolution modelling is essential for translating regional energy transition scenarios into actionable planning insights at local level. For distribution system operators, the approach enables a targeted identification of grid expansion needs at the level of individual substations and across voltage levels, supporting efficient and cost-effective infrastructure planning.

For municipalities, the results provide a robust analytical basis for municipal heat planning by combining technology potential analysis with realistic projections of building-level decisions. The integration of spatially explicit results into interactive tools further enhances the usability of the model outputs for planning and decision-making processes.

Overall, the use case highlights the added value of combining agent-based building simulations with spatially explicit infrastructure modelling, enabling a consistent link between regional scenarios and local implementation planning.
\paragraph{Further details}
Details on the methodology, data processing, and application of the web-based visualisation tool can be found in the corresponding project documentation and online resources \cite{horst_2025_ut, Fischer.2025, Krause.2026, Geiger.04.02.2026, Broadhurst.2026}.
\section{Discussion}
\label{sec:discussion}
As shown in Section~\ref{sec:applications}, \textsc{AgentHomeID} covers a broad range of applications from the national and regional to the municipal scale of building development in Germany. As a descriptive (rather than normative) model, it offers the potential to simulate building stock evolution under different regulatory regimes and cost assumptions in What-if scenarios, thus providing insight into effects of different policies relevant to the political debate. For planning purposes, it takes regional characteristics of single buildings and their owners explicitly into account, rather than working with overall refurbishment rates or building archetypes. The technological side -- envelope and heating system options, design, limitations, costs -- is modelled in great detail, taking the relevant standards into account. Owner heterogeneity in preferences, according to their socio-demographic properties, is explicitly accounted for, thus going beyond individual (or global) cost minimisation. The infrastructure environment of buildings is taken into consideration as well, allowing for the assessment of the interplay between grid development and connected buildings for all infrastructures -- electricity, heat or gas networks. Furthermore, the current German legislative framework is represented in the form of mandatory refurbishment requirements, available subsidies, cost-burden sharing between landlords and tenants, and decarbonisation requirements. The three criteria formulated in the introduction -- flexible, transparent interfaces to local, regional, and national data sets, heterogeneity among building owners and representation of infrastructure and legislative framework -- are thus covered by \textsc{AgentHomeID}.

Beyond covering these three requirements as model features, the applications in Section~\ref{sec:applications} show that each becomes decision-relevant at a different spatial scale.
At the national scale, the differentiated ownership representation and its empirically estimated, class-specific willingness-to-pay produce systematically divergent retrofit uptake and subsidy incidence between owner-occupiers and private landlords -- a distributional pattern that ownership-undifferentiated models cannot reproduce.
At the regional scale, the building-level bottom-up simulation deviates in magnitude and spatial distribution from aggregated top-down projections (e.g.\ NEP scenarios), relocating projected electrification hotspots relevant for grid expansion.
At the urban block scale, the integrated representation of coincident electrified end uses reveals system peak loads that do not coincide with the peaks of individual technologies and would be missed by technology-separated approaches.
Taken together, these results indicate that the distinguishing value of \textsc{AgentHomeID} lies less in any single feature than in spanning ownership differentiation, spatial resolution, and infrastructure coupling within one framework -- a combination not jointly provided by the closest comparators (Tables~\ref{tab:model_comparison_a} and ~\ref{tab:model_comparison_b}).

Further development is possible in several directions: Firstly, as described in Section~\ref{sec:methods}, the temporal evolution on the single building level is determined by various stochastic effects -- e.g.~the status quo of the building, in particular its envelope quality and the age of its heat supply system, is drawn according to a plausible distribution, but at the core randomly inferred.
The changes in occupancy and ownership are similarly drawn, and the ownership type, including heterogeneous preferences, is assigned randomly, again with plausible weighted distributions, but still under uncertainty.
These factors contribute to uncertainty in the evolution of single buildings.
For national studies or larger-scale planning purposes, the different evolutions within an ensemble of buildings can be expected to converge to a plausible and robust overall development at larger scales, even though we cannot expect to predict a single building's evolution accurately.
For planning purposes at smaller scales, such as district heating networks, the uncertainty at the single building level plays a larger role.
To handle such planning tasks, it is necessary to investigate the probability distribution of different realisations of simulated buildings' evolution.
For example, when planning district heating networks, the connection rates are crucial for economic viability.
Instead of evaluating single model runs with respect to connection rates, the probability of a building connecting to the heat grid becomes the more robust measure.
A future direction for model development is thus the investigation and description of ensemble runs, focusing on the heterogeneity of single buildings' evolutions across different realisations.
This ensemble perspective also delimits the level at which the model's outputs should be interpreted: while the framework supports fully disaggregated building-level analyses, including individual owner behaviour and the timing of measures, the robustness of single-building results is inherently limited by the stochastic processes described above, so that interpretability increases with spatial aggregation.
Building-level applications are therefore most informative not as point predictions for individual buildings, but as a means of characterising the robustness, uncertainty, and plausible ranges of development that inform analyses at larger scales.
Such applications are supported by the framework but have not yet been carried out in real-world settings.

Secondly, the model setup as described here neglects agent interactions almost completely, the exception being the competition for overall model budgets (subsidies and renewable fuels). This is in contrast with everyday experience, where learning from other building owners, energy consultants etc.~occurs regularly, in particular when it comes to large investments. \textsc{AgentHomeID} assumes well-informed owners, thus one could argue that energy consulting is implied. Information exchange with other building owners about their experiences with energy-related refurbishment is, however, not included yet.

Thirdly, the foundation of owner agents' choices in empirical surveys has to be updated regularly, since a survey is always conducted under specific circumstances (e.g.~pandemic, energy price shocks, heated political debate about building regulation). Even though preferences have been carefully determined from the empirical data, their validity for future time periods is limited.
\section{Conclusion}
\label{sec:conclusion}
\textsc{AgentHomeID} is a transparent, multi-scale agent-based framework for modelling long-term building stock evolution that explicitly integrates heterogeneous owner behaviour and building properties, techno-economic constraints, infrastructure, and regulatory contexts.
It enables projections under scenarios of changing legislation, including dynamic overall and regional budgets for subsidies and renewable energy carriers.
The framework is designed to support national policy assessment, regional infrastructure planning, and municipal heat planning, while preserving the capability for high-resolution urban applications through its modular, data-flexible architecture.

The model framework is presented in alignment with published reporting guidelines for building stock energy models along with ODD-based sections to facilitate understanding and comparability.
Depending on owner type, agents’ decision-making is based on empirical willingness-to-pay or economic optimisation processes.
Model outputs are data-rich and multi-dimensional, spanning both building- and system-level indicators.
At the building level, the model provides timing and outcomes of refurbishments, primary, final, and useful energy demand and consumption, installed thermal capacities, emissions, investment costs, subsidies, and changes in operating costs.
It also captures distributional aspects by differentiating results by ownership type and income groups.
At the system level, outputs include energy-carrier shares, regionally installed capacities, and projected peak loads, providing indicators essential for regional grid planning and policy evaluation.

Application results on various scales within Germany demonstrate that interactions between policy design and infrastructure deployment strongly shape decarbonisation pace and distributional outcomes of costs and benefits.
Scenarios with binding renewable energy requirements for new heating installations generally accelerate heat-pump uptake and reduce final energy demand, while anticipated gas-grid decommissioning further reinforces electrification and reduces system costs.
Across these scales, the added value of \textsc{AgentHomeID} lies in combining differentiated ownership representation, spatial resolution, and infrastructure coupling within a single framework, a combination not jointly offered by existing building stock models.
Future work will advance uncertainty quantification, further develop the representation of non-residential stocks, pursue open-source dissemination, and further integrate the framework with urban digital twins.
\section*{Data availability}
This work combines publicly available datasets with proprietary regional data.
All input data sources, including providers and access conditions, are documented in Section~\ref{subsubsec:data_sources} (Table~\ref{tab:data_sources}). 
Public data are available from the respective providers.
Due to data protection and confidentiality agreements, detailed building-level data cannot be made publicly available.
Aggregated data and model outputs supporting the findings of this study are available from the authors upon reasonable request.
\section*{Declaration of generative AI and AI-assisted technologies in the writing process}
During the preparation of this work, the authors used generative AI to improve language and readability. After using these tools, the authors reviewed and edited the content as needed and take full responsibility for the publication's content.
\section*{Declaration of competing interest}
The authors declare that they have no known competing financial interests or personal relationships that could have appeared to influence the work reported in this paper.
\section*{CRediT authorship contribution statement}

\textbf{Helen Ganal:} Conceptualization, Methodology, Software, Validation, Investigation, Data curation, Writing -- original draft, Visualization, Project administration, Funding acquisition.
\textbf{Sarah Becker:} Conceptualization, Methodology, Software, Validation, Investigation, Data curation, Writing -- original draft, Visualization, Project administration, Funding acquisition.
\textbf{Sascha Holzhauer:} Conceptualization, Methodology, Software, Validation, Investigation, Data curation, Writing -- original draft, Visualization, Project administration.
\textbf{Thilo Gli\ss{}mann:} Methodology, Software, Validation, Data curation, Writing -- original draft, Project administration.
\textbf{Friedrich Krebs:} Conceptualization, Methodology, Writing -- original draft, Funding acquisition.
\textbf{Philipp H\"artel:} Writing -- review \& editing, Visualization, Supervision.
\textbf{Martin Braun:} Writing -- review \& editing, Supervision.

\medskip
\noindent The model was jointly developed by H.G., S.B., S.H., and T.G.

\section*{Acknowledgements}
The authors gratefully acknowledge funding by the German Federal Ministry for Economy and Climate Protection for the projects DeGeb (FKZ 01LA1808C), OwnPV-Outlook (FKZ 03EI1031A), WAERMER (FKZ 03EI5235A/B), and EnEff:Wärme REBUILT (FKZ 03EN3119A). 
Furthermore, the authors would like to thank Agora Energiewende for employing \textsc{AgentHomeID} in regulatory scenario analyses \citep{Wauer.2025, Wauer.2025c}, and various distribution grid operators for applying it in distribution grid planning.

\FloatBarrier
\bibliographystyle{elsarticle-num} 
\bibliography{cas-refs}
\clearpage
\onecolumn

\appendix
\renewcommand{\thesection}{Appendix \Alph{section}}


\section{Definitions of model comparison criteria}
\label{app:comparison_criteria}

Tables~\ref{tab:comparison_criteria_a} and~\ref{tab:comparison_criteria_b} define the criteria used for the model comparison presented in Tables~\ref{tab:model_comparison_a} and~\ref{tab:model_comparison_b}. The comparison focuses on explicitly documented model capabilities rather than on model labels or intended application domains. The definitions are intended to distinguish features that are represented directly within the model from those that are only represented indirectly, probabilistically, or through external assumptions.


\begin{center}
\captionof{table}{Definitions of the comparison criteria used in Tables~\ref{tab:model_comparison_a} and~\ref{tab:model_comparison_b} (Part a).}
\label{tab:comparison_criteria_a}

\small
\renewcommand{\arraystretch}{1.15}

\begin{tabularx}{\textwidth}{
>{\raggedright\arraybackslash}p{0.34\textwidth}
>{\raggedright\arraybackslash}X}

\toprule
\textbf{Criterion} &
\textbf{Definition}\\
\midrule

Building-level decision units
&
Investment decisions are evaluated for individual buildings (real or representative) rather than exclusively for aggregated building-stock segments.
\\

Decision maker explicitly represented
&
Owners, occupants, or other decision makers are represented with explicitly defined attributes or states instead of being implicitly embedded in buildings or stock segments.
\\

Event-triggered investment decisions
&
Refurbishment or replacement decisions are initiated by explicit events such as component ageing, heating-system failure, ownership change, or regulatory requirements.
\\

Bounded-rational or probabilistic option choice
&
Investment options are selected using stochastic, behavioural, or bounded-rational decision rules rather than purely deterministic optimisation.
\\

Socio-demographic attributes directly affect investment decisions
&
Socio-demographic characteristics (e.g.\ income, age, education) directly influence the investment decision process.
\\

WTP term included in option utility
&
The decision model explicitly accounts for willingness-to-pay or monetised non-financial preferences when evaluating investment options.
\\

WTP informed by empirical decision-maker studies
&
Willingness-to-pay parameters are derived from empirical studies of decision makers rather than solely from calibration or expert assumptions.
\\

Owner-role-specific WTP estimated specifically for the model
&
Owner-specific willingness-to-pay values are estimated for the decision-maker groups represented in the model and directly integrated into the decision process.
\\

Private and institutional owner types
&
The model distinguishes between private households and institutional owners such as companies, cooperatives, or public authorities.
\\

Owner-occupier and landlord differentiation
&
Owner-occupiers and private landlords are represented as separate decision-maker groups.
\\

Ownership-specific decision rules or parameters
&
Different owner groups use different decision rules or evaluation criteria.
\\

Tenant--landlord relationship and cost allocation
&
Rental relationships and the allocation of investment, operating, or carbon costs between tenants and landlords are explicitly represented.
\\

European or multi-country applications
&
The model has been applied consistently across multiple countries using a common modelling framework.
\\

National-scale application
&
The model can represent the complete building stock of a country.
\\

Regional or district-scale application
&
The model supports applications for regions, municipalities, districts, or similar sub-national planning areas.
\\

Neighbourhood or urban-block application
&
The model can be applied to neighbourhoods, urban blocks, or comparable small-scale study areas.
\\

Spatially located building agents
&
Buildings are assigned to explicit spatial locations or spatial units rather than being represented as completely unlocated stock segments.
\\

Explicit real-building and GIS-based input supported
&
Real buildings and GIS-derived building information can be used directly as model input.
\\

Long-term scenario horizon
&
The model simulates long-term building-stock development over several decades.
\\

Annual or multi-year simulation steps
&
Building-stock evolution is represented in recurring discrete simulation steps of one or multiple years.
\\

Endogenous ageing, refurbishment, and stock turnover
&
Component ageing, refurbishment, replacement, demolition, or new construction are determined within the simulation.
\\

\bottomrule
\end{tabularx}
\end{center}

\clearpage


\begin{center}
\captionof{table}{Definitions of the comparison criteria used in Tables~\ref{tab:model_comparison_a} and~\ref{tab:model_comparison_b} (Part b).}
\label{tab:comparison_criteria_b}

\small
\renewcommand{\arraystretch}{1.15}

\begin{tabularx}{\textwidth}{
>{\raggedright\arraybackslash}p{0.34\textwidth}
>{\raggedright\arraybackslash}X}

\toprule
\textbf{Criterion} &
\textbf{Definition}\\
\midrule

Residential building stock
&
Residential buildings are represented explicitly.
\\

Non-residential building stock
&
Non-residential buildings are represented explicitly.
\\

Envelope components represented separately
&
Building envelope components (e.g.\ roof, walls, windows) are modelled individually.
\\

Multiple heating technologies and energy carriers
&
The model distinguishes several heating technologies and energy carriers.
\\

Coupled envelope and heating-system choices
&
Envelope refurbishment and heating-system replacement are evaluated jointly or through interacting decision processes.
\\

Spatially differentiated infrastructure availability
&
Infrastructure availability varies spatially according to location or settlement characteristics.
\\

Building-level infrastructure connection status
&
Infrastructure availability is determined individually for each modelled building.
\\

Building-specific renewable-heat feasibility
&
Building- or site-specific suitability for renewable heating technologies is explicitly considered.
\\

Explicit network topology or capacity representation
&
Network topology, connection structure, or infrastructure capacities are represented explicitly.
\\

Infrastructure evolution and municipal heat-planning integration
&
Temporal infrastructure development and/or municipal heat-planning information are represented.
\\

Carbon pricing represented in operating costs
&
Carbon pricing directly affects operating costs.
\\

Investment subsidies or operating support
&
Financial support schemes influence investment decisions.
\\

Technology bans, requirements, or efficiency standards
&
Technology restrictions or minimum efficiency requirements influence feasible investment options.
\\

Detailed country-specific legal interactions
&
Multiple interacting legal and regulatory provisions of a specific national framework are represented.
\\

Finite policy budgets or resource constraints
&
Policy budgets or limited infrastructure or fuel resources constrain model outcomes.
\\

Useful heat demand per modelled building unit
&
Useful space-heating demand is calculated individually for each modelled building unit.
\\

Final energy demand per modelled building unit
&
Final energy demand is calculated individually, accounting for heating-system efficiencies and energy carriers.
\\

Norm-based design heating load
&
Heating-system sizing is based on a normative design heating-load calculation.
\\

Hourly demand profiles generated endogenously
&
The model generates hourly demand or load profiles directly during the simulation.
\\

Post-processed load profiles or peak-load indicators
&
Hourly loads or peak demands are derived afterwards from model outputs using external load-profile approaches.
\\

GHG emissions calculated per modelled building unit
&
Greenhouse-gas emissions are determined individually for each modelled building unit.
\\

Investment and operating costs per decision unit
&
Investment and operating costs are calculated individually for each building or decision maker.
\\

Distributional results by ownership or income group
&
Model outputs can be evaluated separately for different ownership or socio-economic groups.
\\

\bottomrule
\end{tabularx}
\end{center}

\clearpage

\section{Data differentiation in AgentHomeID}
\label{app:data_differentiation}

Table~\ref{tab:data_sources} provides an overview of the principal input data represented in AgentHomeID, including the level of differentiation, data sources, and implementation basis for each model component.

\begin{center}
    \captionof{table}{Differentiation of the principal input data represented in AgentHomeID. The table reflects the implemented data model; individual studies may replace national distributions and scenario pathways with regional inputs.}
    \scriptsize
    \setlength{\tabcolsep}{3pt}
    \renewcommand{\arraystretch}{1.08}
    \begin{tabularx}{\textwidth}{@{}>{\raggedright\arraybackslash}p{0.14\textwidth}>{\raggedright\arraybackslash}p{0.23\textwidth}>{\raggedright\arraybackslash}X>{\raggedright\arraybackslash}p{0.24\textwidth}@{}}
        \toprule
        Aspect & Attributes & Differentiation in AgentHomeID & Data source / implementation basis \\
        \midrule
        Spatial representation
        & Building location, stock weight and model scale
        & National: Weighted unlocalised sample of representative buildings; local/regional stock: localised building-level data
        & National: German Microcensus SUF \cite{SUF}; local/regional: census and regional geodata \cite{StatistischesBundesamt.2022} using UrbanTwin data \citep{horst_2025_ut} \\

        Residential typology and geometry
        & Type, construction year, dwellings, usable/rental area, volume and envelope areas
        & Four types (detached/semi-detached, terraced, small and large multi-family); five construction-age classes
        & National: SUF and Census; typology based on IWU residential typology \cite{SUF,iwu_tabula}, type building geometries inferred from national UrbanTwin data \citep{horst_2025_ut}; localised: UrbanTwin \citep{horst_2025_ut} \\

        Non-residential stock
        & Function \& age class, geometry, ownership and initial technical state
        & Nine functional types and three construction-age classes; building-level scaling for local stocks
        & National: ENOB:dataNWG and German non-residential typology \cite{enob_data_nwg,Horner.2022} or localised: UrbanTwin \citep{horst_2025_ut} \\

        Building envelope
        & Ground/lower closure, walls, roof and windows; component age, lifetime and thermal transmittance
        & State and refurbishment level tracked separately for each component; stochastic lifetime and replacement year
        & IWU stock survey and component lifetimes \cite{iwu:2016,bte}; calibration of lifespans \\

        Owner and household
        & Owner type, age, income, education, gender, wealth, occupancy and move-in year
        & Owner-occupier, private landlord, owners' association, corporate, public and cooperative owner
        & SUF and Census distributions \cite{SUF,StatistischesBundesamt.2022}; UrbanTwin \citep{horst_2025_ut} \\

        Heating supply
        & Heating system, energy carrier, distribution type, system age and lifetime
        & Boilers (oil, gas or hydrogen; with/without solar), air/ground-source heat pumps, biomass and district/local heat
        & Initial stock from SUF \citep{SUF} or UrbanTwin \citep{horst_2025_ut}; technology-specific performance, lifetime \\

        Energy infrastructure
        & Availability of gas/hydrogen grids, district/local heat and heat pumps; gas-grid decommissioning
        & Building- and year-specific availability; Local information for regional runs if available, otherwise probability distributions
        & Census/regional infrastructure data and scenario tables; municipal heat plans \cite{StatistischesBundesamt.2022,WPG} \\

        Decision heterogeneity
        & Latent class, willingness-to-pay, budget and option attributes
        & Random utility for private owners and owners' associations; net present value for institutional owners; trigger-specific cancellation; latent class attribution depending on owner type and based on socio-demographic parameters
        & Econometric latent-class/mixed-logit input tables; owner-specific financial assumptions \cite{Bender2026} \\

        Stock dynamics
        & New construction, demolition, owner/tenant change and component failure
        & Annual, region-specific new-build rates; stochastic component lifetimes and event triggers
        & Census/official stock statistics and configured scenario parameters \cite{StatistischesBundesamt.2022,bte} \\

        Costs and environmental factors
        & Investment, operating and fuel costs; subsidies; $\mathrm{CO_2}$ and primary-energy factors
        & Technology, building state, owner and year; price/emission paths, price indices and learning rates
        & IWU cost data \cite{Hinz.10.08.2015}; official indices \citep{koch_2021} and scenario-specific model input tables \\

        Policy and resources
        & Efficiency requirements, technology restrictions, subsidies and finite subsidy/fuel budgets
        & Regulation- and year-specific schedules
        & GEG, BEG, WPG, BEHG/EU ETS 2 and scenario configurations/hypothetical regulation scenarios \cite{GEG,BEG,WPG,BEHG,ETS2} \\
        \bottomrule
    \end{tabularx}
    \label{tab:data_sources}
\end{center}

\clearpage
\onecolumn

\section{NPV assumptions and decision parameters}
\label{app:npv}

Table~\ref{tab:npv_assumptions} summarises the main parameter assumptions
used in the NPV-based investment assessment of institutional owners.
For self-use-oriented assessments, applied to cooperatives and selected
public authorities, avoided fuel-cost savings are evaluated over the
technical lifetime of the respective building component or heating system.
For rental-oriented assessments, applied to private property companies
and selected public authorities, the evaluation horizon is defined by an
owner-specific maximum payback period. In both approaches, owner- and
trigger-specific rates of return are used as discount rates. The table
further reports the main rental-property assumptions affecting the
recoverable investment costs.

\begin{center}
\captionof{table}{Main parameter assumptions used in the NPV-based investment assessment of institutional owners.}
\label{tab:npv_assumptions}

\small
\renewcommand{\arraystretch}{1.10}

\begin{tabularx}{\textwidth}{
    >{\raggedright\arraybackslash}p{4.2cm}
    >{\raggedright\arraybackslash}X
    >{\raggedright\arraybackslash}p{3.2cm}
}
\toprule
\textbf{Parameter} &
\textbf{Differentiation} &
\textbf{Value} \\
\midrule

\multicolumn{3}{l}{\textit{Economic decision parameters}} \\

Rate of return
(used as discount rate)
& Private property companies: necessary / optional refurbishment
& 3\% / 9\% \\

&
Cooperatives: necessary / optional refurbishment
& 1\% / 2\% \\

&
Public authorities: necessary / optional refurbishment
& 1\% / 4\% \\

Maximum payback period
& Private property companies; rental-oriented NPV
& 10 years \\

&
Public authorities; rental-oriented NPV
& 20 years \\

\midrule
\multicolumn{3}{l}{\textit{Evaluation horizons for self-use-oriented NPV}} \\

Mean technical lifetime
& Lower building closure
& 100 years \\

&
Exterior walls
& 80 years \\

&
Roof
& 80 years \\

&
Windows
& 40 years \\

&
Condensing boiler, with / without solar thermal support
& 18 years \\

&
Low-temperature boiler, with / without solar thermal support
& 21 years \\

&
Constant-temperature boiler, with / without solar thermal support
& 24 years \\

&
Pellet heating, with / without solar thermal support
& 15 years \\

&
Air-source heat pump, with / without ventilation
& 18 years \\

&
Brine-source heat pump, with / without ventilation
& 20 years \\

&
District heating
& 20 years \\

\midrule
\multicolumn{3}{l}{\textit{Rental-property assumptions}} \\

General modernisation levy
& Standard / tenant-market case
& 8\% / 4\% \\

Share of general rent-increase cap
& Share available for energy-related refurbishment
& 50\% \\

Modernisation-related
rent-increase cap
& Existing rent below EUR 7/(m$^2$\,month)
& EUR 2/(m$^2$\,month) \\

&
Existing rent from EUR 7/(m$^2$\,month)
& EUR 3/(m$^2$\,month) \\

Heating-system
modernisation levy
& Without subsidies / with subsidies
& 8\% / 10\% \\

Heating-system
rent-increase cap
& Newly installed heating system
& EUR 0.50/(m$^2$\,month) \\

Reference-rent increase
& Tenant-market / other market assumption
& 5\% / 15\% over 3 years \\

\bottomrule
\end{tabularx}
\end{center}

\clearpage
\twocolumn

\section{Envelope refurbishment options and efficiency levels}
\label{app:env_opts}

Table~\ref{tab:env_opts} summarises the refurbishment options and efficiency
levels considered for the building envelope components within the model.

\begin{center}
\captionof{table}{List of predefined envelope target options. \texttt{\old} indicates the component remains unrefurbished; the refurbishment levels are: \texttt{\slev{1}} -- minimum legal requirement in case of refurbishment of the respective component, \texttt{\slev{2}} -- minimum condition to receive subsidies, \texttt{\slev{3}} -- highest efficiency standard short of passive house standard.}
    \label{tab:env_opts}
\begin{tabularx}{\linewidth}{ZZZZZ}
\toprule
 Option ID & Lower building closure & Outer walls & Roof & Windows \\
\midrule
0 & \old & \old & \old & \old \\
1 & \old & \old & \old & \slev{3} \\
2 & \old & \old & \slev{1} & \old \\
3 & \old & \old & \slev{2} & \old \\
4 & \old & \old & \slev{3} & \old \\
5 & \old & \slev{1} & \old & \slev{3} \\
6 & \old & \slev{2} & \old & \slev{3} \\
7 & \old & \slev{3} & \old & \slev{3} \\
8 & \old & \old & \slev{1} & \slev{3} \\
9 & \old & \old & \slev{2} & \slev{3} \\
10 & \old & \old & \slev{3} & \slev{3} \\
11 & \slev{1} & \old & \old & \slev{3} \\
12 & \slev{2} & \old & \old & \slev{3} \\
13 & \slev{3} & \old & \old & \slev{3} \\
14 & \slev{1} & \old & \slev{1} & \old \\
15 & \slev{2} & \old & \slev{2} & \old \\
16 & \slev{3} & \old & \slev{3} & \old \\
17 & \old & \slev{1} & \slev{1} & \slev{3} \\
18 & \old & \slev{1} & \slev{2} & \slev{3} \\
19 & \old & \slev{1} & \slev{3} & \slev{3} \\
20 & \old & \slev{2} & \slev{1} & \slev{3} \\
21 & \old & \slev{2} & \slev{2} & \slev{3} \\
22 & \old & \slev{2} & \slev{3} & \slev{3} \\
23 & \old & \slev{3} & \slev{1} & \slev{3} \\
24 & \old & \slev{3} & \slev{2} & \slev{3} \\
25 & \old & \slev{3} & \slev{3} & \slev{3} \\
26 & \slev{1} & \slev{1} & \slev{1} & \slev{3} \\
27 & \slev{1} & \slev{1} & \slev{2} & \slev{3} \\
28 & \slev{1} & \slev{1} & \slev{3} & \slev{3} \\
29 & \slev{1} & \slev{2} & \slev{1} & \slev{3} \\
30 & \slev{2} & \slev{2} & \slev{2} & \slev{3} \\
31 & \slev{2} & \slev{2} & \slev{3} & \slev{3} \\
32 & \slev{1} & \slev{3} & \slev{1} & \slev{3} \\
33 & \slev{2} & \slev{3} & \slev{2} & \slev{3} \\
34 & \slev{3} & \slev{3} & \slev{3} & \slev{3} \\
\bottomrule
\end{tabularx}
\end{center}

\end{document}